\documentclass[%
 aip,
 amsmath,amssymb,
 reprint,%
]{revtex4-1}
\usepackage[percent]{overpic}
\usepackage{subcaption} % in the preamble
\usepackage{amsmath}    % for math formatting
\usepackage{booktabs}   % for nicer tables
\usepackage{array}   % for >{\hspace} adjustments
\newcolumntype{C}[1]{>{\centering\arraybackslash}p{#1}}
\usepackage{multirow}   % for multi-row cells
\usepackage{siunitx}    % optional, for number alignment
\usepackage{comment}

\usepackage{graphicx}% Include figure files
\usepackage{dcolumn}% Align table columns on decimal point
\usepackage{bm}% b\begin{figure*}
\usepackage{xcolor}
\usepackage{amsmath}    % For math equations
\usepackage{overpic}    % To write text over images
\usepackage{xparse}
\NewDocumentCommand{\panellabel}{mO{white}}{\textcolor{#2}{\large\textbf{(#1)}}}

\usepackage[utf8]{inputenc}
\usepackage[T1]{fontenc}
\usepackage{mathptmx}
\usepackage{etoolbox}

\usepackage[english]{babel}
\makeatletter
\addto\captionsenglish{\def\l@en{\l@english}}
\makeatother

\usepackage[normalem]{ulem}
\usepackage{xcolor}
\usepackage{hyperref}
\definecolor{myred}{HTML}{B55A43}
\definecolor{mygreen}{HTML}{3D8C40}

\definecolor{changes}{rgb}{0.1,0.1,1}
\newcolumntype{C}[1]{>{\centering\arraybackslash}p{#1}}

\makeatletter
\def\@email#1#2{%
 \endgroup
 \patchcmd{\titleblock@produce}
  {\frontmatter@RRAPformat}
  {\frontmatter@RRAPformat{\produce@RRAP{*#1\href{mailto:#2}{#2}}}\frontmatter@RRAPformat}
  {}{}
}%

\makeatother
\begin{document}

\preprint{AIP/123-QED}

\title[Electron neural closure for turbulent magnetosheath simulations: energy channels]{Electron neural closure for turbulent magnetosheath simulations: energy channels}
% Force line breaks with \\
\author{G. Miloshevich}
%\homepage{georgemilosh.github.io}
 \email{george.miloshevich@kuleuven.be}
\affiliation{Centre for mathematical Plasma Astrophysics, Department of Mathematics, KU Leuven, Celestijnenlaan 200B, 3001 Leuven, Belgium}

\author{L. Vranckx}%
\affiliation{Centre for mathematical Plasma Astrophysics, Department of Mathematics, KU Leuven, Celestijnenlaan 200B, 3001 Leuven, Belgium}

\author{F. N. de Oliveira Lopes}
\affiliation{Centre for mathematical Plasma Astrophysics, Department of Mathematics, KU Leuven, Celestijnenlaan 200B, 3001 Leuven, Belgium}

\author{P. Dazzi}
\affiliation{Centre for mathematical Plasma Astrophysics, Department of Mathematics, KU Leuven, Celestijnenlaan 200B, 3001 Leuven, Belgium}

\author{G. Arrò}
\affiliation{Department of Physics, University of Wisconsin-Madison, Madison, WI 53706, USA}%

\author{G. Lapenta$^{\dagger}$}
 \thanks{$^{\dagger}$Deceased, May 2024.}
\affiliation{Centre for mathematical Plasma Astrophysics, Department of Mathematics, KU Leuven, Celestijnenlaan 200B, 3001 Leuven, Belgium}

\date{\today}% It is always \today, today,

\begin{abstract}
In this work, we introduce a non-local five-moment electron pressure tensor closure parametrized by a Fully Convolutional Neural Network (FCNN). Electron pressure plays an important role in generalized Ohm's law, competing with electron inertia. This model is used in the development of a surrogate model for a fully kinetic energy-conserving semi-implicit Particle-in-Cell simulation of decaying magnetosheath turbulence. We achieve this by training FCNN on a representative set of simulations with a smaller number of particles per cell and showing that our results generalise to a simulation with a large number of particles per cell. We evaluate the statistical properties of the learned equation of state, with a focus on pressure-strain interaction, which is crucial for understanding energy channels in turbulent plasmas. The resulting equation of state learned via FCNN significantly outperforms local closures, such as those learned by Multi-Layer Perceptron (MLP) or double adiabatic expressions. We report that the overall spatial distribution of pressure-strain and its conditional averages are reconstructed well. However, some small-scale features are missed, especially for the off-diagonal components of the pressure tensor. Nevertheless, the results are substantially improved with more training data, indicating favorable scaling and potential for improvement, which will be addressed in future work. 

\end{abstract}

\maketitle

\section{Introduction}\label{sec:intro} 
Understanding and predicting the behavior of collisionless space plasmas in the near-Earth environment presents a fundamental scientific challenge due to its multi-scale nature. Many of the important theoretical and numerical results can be tested in this environment. Of particular interest are the energy exchanges and dissipation in collisionless plasmas, driven by reconnection and wave-particle interactions. Turbulence drives the formation of thin current sheets that reconnect and may drive secondary reconnection processes. Thus, these phenomena are intimately linked in plasmas~\cite{stawarz_turbulence-driven_2022}. Figuring out how plasma is energized in such turbulent environments was highlighted as one of the critical research questions for future space missions~\cite{retino_particle_2022}. These processes create extreme events posing serious risks to our infrastructure~\cite{thira_2019_2019}. 

The Earth's magnetosheath, which separates the bow shock from the magnetopause, is located at the interface between the Earth's magnetosphere system and the solar wind that drives space weather. The {interplay between turbulence and reconnection}~\cite{lazarian_turbulent_2015,lazarian_3d_2020} in the magnetosheath has recently been a subject of study, for instance, the influence of turbulence on the length of current sheets formed~\cite{stawarz_interplay_2024}. This leaves open questions, such as understanding the interaction between magnetic reconnection and other types of {waves and instabilities}~\cite{nakamura_outstanding_2025,khotyaintsev_collisionless_2019} that contribute to plasma heating.
Indeed, the magnetosheath is also home to a variety of modes that are driven by pressure anisotropies, such as whistler waves, which have been recently associated with driving the formation of electron magnetic holes~\cite{arro_generation_2023,espinoza-troni_electron_2025}. These processes were studied using ECsim~\cite{lapenta_exactly_2017} (an energy-conserving semi-implicit Particle-in-Cell (PIC) code), which allows moderately large domains spanning {ion Larmor radius-electron inertial length scales} and mass ratios $m_i/m_e \sim 100$. 

{PIC codes}~\cite{fonseca_osiris_2002,fonseca_one--one_2008,germaschewski_plasma_2016} are more efficient than higher fidelity Vlasov fully kinetic simulations~\cite{allmann-rahn_muphyii_2024}. Such simulations can only be performed for certain parameter ranges and spatial scales. A more efficient approach involves Reduced Order Models (ROMs), such as hybrid Vlasov~\cite{valentini_hybrid-vlasov_2007,palmroth_vlasov_2018} and hybrid PIC codes~\cite{behar_menura_2022} and multi-fluid codes~\cite{dong_global_2019,wang_global_2022} which have been developed for global modelling of magnetospheres. {By \emph{hybrid} models we refer to models which have kinetic ions and fluid electrons}. These models have made significant advances in our ability to represent the physics at mesoscales, typically covering MHD and sub-ion scales, but crucially omitting electron scales. Electrons are typically assumed to be polytropic or even isothermal. This is in stark contrast to the physics which takes place at microscopic scales, whereby electrons are de-magnetized in the process of collisionless magnetic reconnection~\cite{birn_geospace_2001}, where electrons also play a key role in modifying dynamics~\cite{hesse_magnetic_2020}. These processes contribute to the heating of electrons that is not captured in hybrid models, as confirmed by simulations of the Hermean magnetosphere carried out using fully kinetic simulations~\cite{lapenta_we_2022} when compared to much cooler electrons found in hybrid simulations~\cite{travnicek_mercurys_2010}. Thus, in order to understand and predict energization of plasma, it is important to couple high-fidelity fully kinetic simulations with ROMs to properly represent the physics of both electrons and ions.

One of the potentially promising avenues involves coupling PIC codes to fluid codes in an attempt to model complex objects, such as Earth's magnetosphere, more accurately~\cite{daldorff_two-way_2014,toth_extended_2016,wang_simulation_2022,chen_fleks_2023}. This allows resolving the magnetotail with a PIC simulation in a small box coupled to a larger fluid magnetospheric simulation. {This approach is not justified if the entire box is dominated by kinetic physics~\cite{li_numerical_2023} and fully kinetic modelling should be considered in this case}.

An alternative approach involves selecting an appropriate closure relation for electrons that is embedded within hybrid simulations. The problem of collisionless fluid closure has a long history in plasma physics, dating back to the works of \citet{chew_boltzmann_1997}, which postulated simple double-adiabatic relationships that later became known as the CGL equations, named after the authors Chew, Goldberger, and Low. It is well known that such closures break down under moderate values of Finite Larmor Radius (FLR) effects due to the thermal gyration of particles around magnetic fields. A more advanced version of such anisotropic pressure closure has been proposed in a series of seminal works~\cite{le_equations_2009,ohia_demonstration_2012}, which proposes interpolation between the trapped particle dynamics, more closely resembling the CGL-like dynamics, and passing particle dynamics, which corresponds to the Boltzmann limit where the plasma is isothermal along the field lines. The evidence for this behavior was found~\cite{wetherton_validation_2019} in Magnetospheric Multiscale (MMS~\citet{burch_electron-scale_2016}) observations. One of the limitations of CGL-like models is the inability to deal with \emph{agyrotropy}, unequal dispersions of the Velocity Distribution Function (VDF) perpendicular to the local field. Enhanced agyrotropy is typically present near the current sheets on the scale of inertial length. Early works~\cite{hesse_role_2004} have shown the importance of such features of the pressure tensor in the total electric field in guide field magnetic reconnection.  

It is possible to close the fluid equations by providing an equation of state for heat flux, rather than pressure, in which case pressure is modeled dynamically. This allows processes such as Landau damping to be represented, which is typically not possible within a fluid framework. Such models are usually referred to as \emph{Landau fluids}, and were introduced by~\citet{hammett_fluid_1990} and later developed by~\citet{passot_collisionless_2007}. Crucially, they operate under linear response approximation. Under large to moderate guide fields, such models do indeed capture the main features of magnetic reconnection~\cite{finelli_bridging_2021}, with some exceptions such as electron-cyclotron instability. Importantly, it is challenging to apply them in low guide fields~\cite{wang_comparison_2015}. Although some progress has been achieved in later works~\cite{ng_simulations_2017}, the strong system-size dependence of the average reconnection rate observed in kinetic and hybrid simulations~\cite{stanier_role_2015} was not completely reproduced, leaving room for new developments. 

More recently, Machine Learning (ML) approaches have been widely applied in fluid dynamics~\cite{vinuesa_enhancing_2022} and the geophysical sciences, including plasma physics~\cite{anirudh_2022_2022}. In Earth weather, neural subgrid closures have been successfully implemented and currently rival fully physics-based solvers~\cite{kochkov_neural_2024}. A \emph{subgrid closure} is one where small-scale processes are parametrized using approximations or a statistical model. This is a popular approach in Large Eddy Simulations (LES) with deterministic or stochastic parametrizations developed using Variational Autoencoders (VAEs) and Generative Adversarial Networks (GANs)~\cite{perezhogin_generative_2023}. While electron pressure or heat flux closure is not equivalent to this, there are parallels that have already been explored using a relatively simple ML architecture in the case of the GEM challenge, i.e., modelling collisionless magnetic reconnection. \citet{laperre_identification_2022} has applied Histogram Gradient Boosting Regressor (HGBR) and Multi Layer Perceptron (MLP) for this task to map lower order moments such as density, velocity, electric and magnetic field to pressure tensor and heat flux vector. They have reported difficulties reconstructing off-diagonal components related to agyrotropy, especially in the region bound by the reconnection X-lines. 

{Fluid closure in plasma has already been addressed using Convolutional Neural Networks (CNNs), which are the backbone of image analysis in computer vision. In particular, a one-dimensional Hammett-Perkins closure was previously sought~\cite {ma_machine_2020}; however, a single dense layer with a Discrete Fourier Transform (DFT) appeared to perform better. The authors argued that this was due to the availability of an analytic solution, which would not be the case in higher dimensions. CNNs were also trained to represent an analytic form of collisional Braginskii’s and Hammett-Perkins closure given a range of macroscopic input profiles~\cite{maulik_neural_2020}. More complex architectures, such as VNet, have been applied to nonlocal Vlasov-Poisson with Bhatnagar, Gross, and Krook collision operator~\cite{bois_neural_2022}. This data-driven closure was reported to be numerically stable but costly, with potential advantages in higher dimensions. More recently, a one-dimensional Landau fluid closure was obtained using a Fourier Neural Operator (FNO) and reportedly outperformed the Hammett-Perkins closure~\cite{huang_machine-learning_2025}. }

In parallel, there have been approaches focusing on symbolic and sparse regression~\cite{camps-valls_discovering_2023}, where equations are extracted directly from simulation or observational data. For instance,~\citet{alves_data-driven_2022} have extracted polytropic closures from collisionless shock simulations using a popular method referred to as SINDy (Symbolic Identification of Nonlinear Dynamics)~\cite{brunton_discovering_2016}. This method has been subsequently applied to extract heat flux closures in magnetic reconnection~\cite{donaghy_search_2023} and some electrostatic phenomena~\cite{ingelsten_data-driven_2025}. While we believe that the approach is promising and there are many unexplored avenues in symbolic regression in general~\cite{landajuela_unified_2022}, this falls beyond the scope of this manuscript. In addition, there are some disadvantages to using SINDy, such as the need to choose a specific library of terms, i.e., a preselected choice of potential expressions, and limitations on expressivity due to polynomial representation. 

Therefore, our goal is to learn pressure and a heat flux closure using neural networks trained on multiple simulations performed with the fully kinetic code ECsim~\cite{lapenta_exactly_2017}. Our approach differs from~\citet{laperre_identification_2022} in that we consider a problem of magnetosheath turbulence, which is a significantly more challenging setting due to the presence of chaotic turbulent eddies. To improve upon the state of the art, we introduce a Fully Convolutional Neural Network (FCNN) that operates on patches rather than locally and study the importance of using global closure for both pressure and heat flux. We evaluate the performance of the new closure by inspecting energy channels using a scale filtering approach and discussing both the spatial structures and overall statistics. This includes evaluating anisotropy versus $\beta_\|$ plots that are well bounded by whistler and electron firehose instabilities in our simulations, as well as the learned closure. 

The manuscript is organized as follows. In section~\ref{sec:methodology}, we discuss the methodology starting from the description of the dataset in section~\ref{ref:sim-data}, then in section~\ref{sec:adiabatic}, we introduce the closure theory and simple adiabatic closures like CGL {, also introducing pressure-strain and related quantities. In section~\ref{sec:results} we discuss results consisting of neural pressure and heat flux closure in section~\ref{sec:pressure_heat}. Next, we characterise predicted anisotropies in section~\ref{sec:anisotropies}, and evaluate the energy channels in section~\ref{sec:channel_evaluations}. We address the issue of generalization across low and high fidelity PIC simulations in section~\ref{sec:ablation}. This is followed by a discussion in section~\ref{sec:discussion} and a conclusion in section~\ref{sec:conclusion}}.

\begin{figure*}
\includegraphics[width=1.0\linewidth]{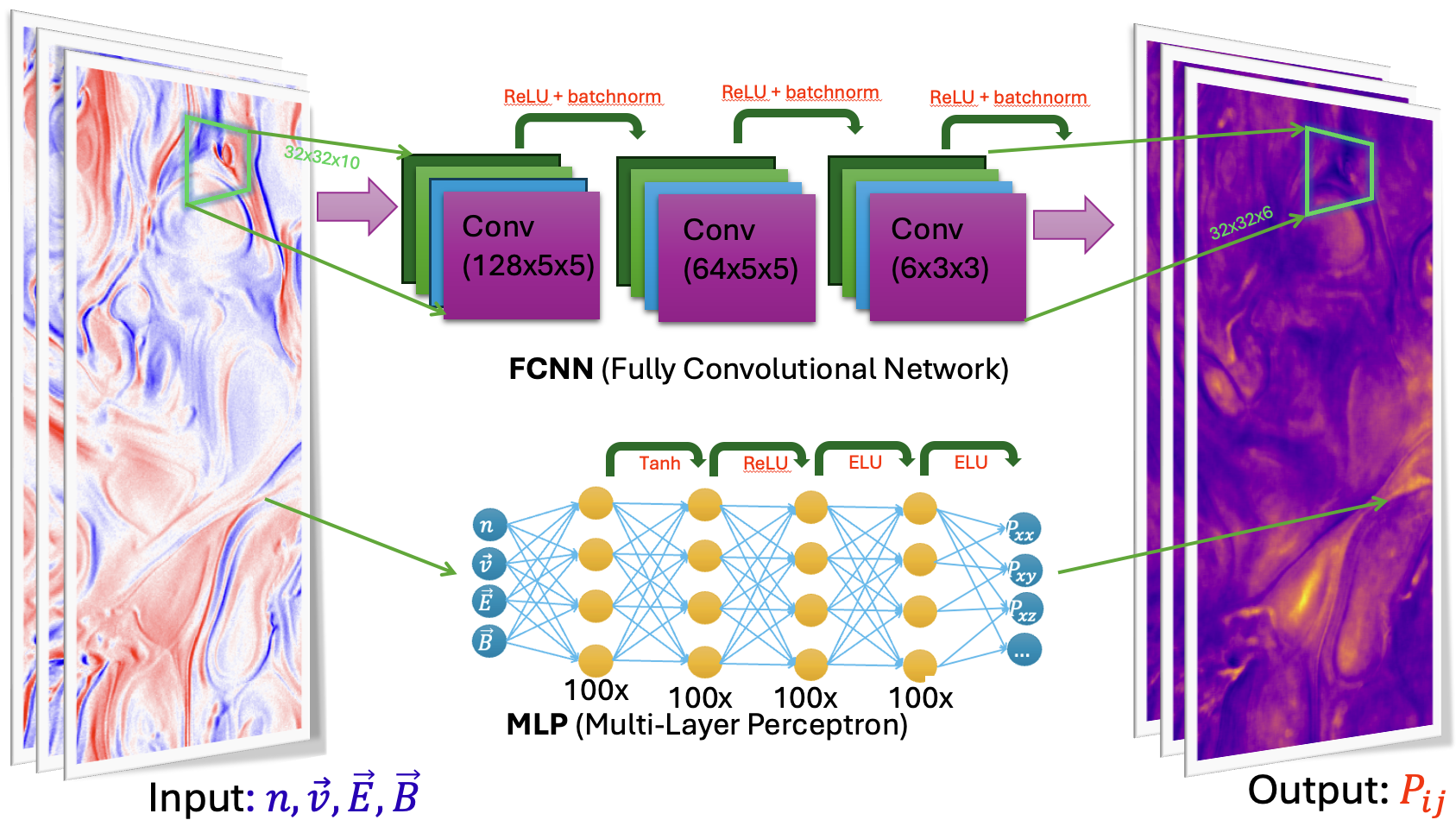}% Here is how to import EPS art
%'/Users/u0167590/Documents/2025 - 2026 - Leuven/Offlineclosurepaper/Thierry.pptx'
%/volume1/scratch/georgem/closure/models/peppe/sigma0_haydn/FCNN/P/PiD.ipynb
\caption{\label{fig:architecture} The input/output and architecture of the Fully Convolutional Neural Network (FCNN), top, and Multi Layer Perceptron (MLP), bottom. The images on the left are being fed into both architectures using either a patch-based approach, as indicated by a green frame in the case of FCNN, or a point-based approach, as indicated by a green arrow in the case of MLP. On the right, the output pressure tensor is plotted. The internal architecture of FCNN consists of 3 convolutional layers with (number of channels, kernel dimension 1, kernel dimension 2) indicated on top. The green arrows indicate the application of activation functions and batch normalization. MLP consists of 4 layers, each with 100 neurons.}
\end{figure*}

%%%%%%%%%%%%%%%%%%%%%%%%%%%%%%%%%%%%%%%

%%%%%%%%%%%%%%%%%%%%%%%%%%%%%%%%%%%%%%%

\section{Methodology}\label{sec:methodology}
Collisionless plasmas evolve according to the Vlasov equation, which can be written for each species
\begin{equation}\label{eq:vlasov}
\frac{\partial f_s}{\partial t}+\mathbf{v} \cdot \frac{\partial f_s}{\partial \mathbf{x}}+\frac{e_s}{m_s}\left(\mathbf{E}+\frac{\mathbf{v} \times \mathbf{B}}{c}\right) \cdot \frac{\partial f_s}{\partial \mathbf{v}}=0,
\end{equation}
where $f_s = f_s(\mathbf{x},\mathbf{v},t)$ stands for one-particle distribution function {at position $\mathbf{x}$ having velocity $\mathbf{v}$} for species $s$, while $\mathbf{E}$ and $\mathbf{B}$ are self-consistent electromagnetic fields and {$c$ is speed of light}. The Vlasov equation is solved coupled to Maxwell's equations
\begin{equation}\label{eq:faraday}
\frac{\partial \mathbf{B}}{\partial t} = -c \nabla \times \mathbf{E}
\end{equation}
and
\begin{equation}\label{eq:maxwell}
\nabla \times \mathbf{B}=\frac{1}{c}\left(4 \pi \mathbf{J}+\frac{\partial \mathbf{E}}{\partial t}\right)
\end{equation}
Here $\mathbf{J}$ stands for the total current, defined as
\begin{equation}
    {\mathbf{J} = \sum_s e_s n_s \mathbf{v}_s ,}
\end{equation} where $n_s$ corresponds to the bulk density.
This deceptively simple-looking system of equations can be solved numerically using Vlasov~\cite{allmann-rahn_muphyii_2024} or Particle-in-Cell codes~\cite{markidis_energy_2011}; however, due to the multiscale nature of plasma, it becomes prohibitively expensive for certain problems. 

\subsection{Simulation data}\label{ref:sim-data}

In this study, we use a 2D-3V simulation described in \citet{arro_spectral_2022} (hereafter referred to as run A) that was carried out using the energy-conserving semi-implicit Particle-in-Cell code ECsim \cite{lapenta_exactly_2017}, under conditions comparable to those of Earth's magnetosheath. The main parameters of the run are tabulated in column A of Table~\ref{tab:sim_params}. 

{For run A we consider a 2D square periodic box of length $L\!=\!64\,d_i$, sampled by a uniform grid with $2048^2$ points. Here $d_i = c/\omega_{pi}$ stands for the ion inertial length and $\omega_{pi} = \sqrt{4 \pi e_i n_i^2/m_i} $ the ion plasma frequency. We employ $5000$ particles per cell for both ions and electrons, with mass ratio $m_i/m_e\!=\!100$ (with $m_i$ and $m_e$ being ion and electron masses). Particles are initialized from a Maxwellian distribution with uniform density, uniform temperature, and plasma beta (defined as $\beta_s = 8\pi n_s k_B T_s/B^2$) equal to $\beta_i\!=\!8$ for ions and $\beta_e\!=\!2$ for electrons. A uniform initial magnetic field $\textbf{B}_0\!=\!B_0\widehat{\textbf{z}}$ is present, and turbulence is induced by random magnetic field and velocity fluctuations with wavenumbers $k$ in the range $1\leqslant k/k_0\leqslant 4$ (with $k_0\!=\!2\pi/L$). Magnetic field and velocity fluctuations have amplitudes $\delta B_{rms}/B_0\!\simeq\!0.9$ and $\delta u/c_A\!\simeq\!3.6$ (where $c_A$ is the Alfvén speed referred to $B_0$). The time step used to evolve the simulation is $\Delta t\!=\!0.05\,\Omega_e^{-1}$ (with $\Omega_e = e_e B_0/m_e c$ being the electron cyclotron frequency). Time is normalized to $\Omega_e^{-1}$ and the ratio of the plasma frequency to the cyclotron frequency is $\omega_{p,i}/\Omega_i = 100$ for the ions and $\omega_{p,e}/\Omega_e = 10$ for the electrons. Additional information is provided in \citet{arro_spectral_2022,arro_generation_2023,espinoza-troni_electron_2025}.}  

%As will become clear from the section~\ref{sec:results}, a single simulation does not produce satisfactory results when supervised learning is applied. Furthermore, if only a single run is given, the only possible data splits are random or chronological, which we do not consider robust tests for closure. 

We have performed six supporting simulations (referred to as run B) that are initialized using very similar values of the parameters, except for $\delta B/B$, which is larger in run {A} from the start of the simulation and remains larger over the whole simulation time {(see Table~\ref{tab:sim_params})}. {Each member of B starts from a new random realization of the magnetic and velocity field fluctuations defined above with the prescribed levels of $\delta B/B$}. Due to computational costs, we limited the six runs to 256 particles per cell; therefore, B runs tend to be noisier. For these reasons, we can use {A as a higher-fidelity reference point.} %as a generalization test when applying closure trained (section~\ref{sec:neural_closure}) on runs B. 

\begin{table}
\caption{\label{tab:sim_params} Initial simulation parameters in the run A~\cite{arro_spectral_2022}, which has 5000 particles per cell, vs. runs B (of which there are 6), which have 256 particles per cell.    }
\begin{ruledtabular}
\begin{tabular}{l|cr}
Simulation& A: 5000 ppcell %\footnote{Note b.}
& B: 256 ppcell\\
\hline
$\delta B/ B$ & 0.71 & 0.59\\
$\delta V_i/ V_A$ & 1.1 & 1.1\\
$\delta V_e/ V_A$ & 1.1 & 1.1\\
$\beta_i$ & 5.3 & 5.7\\
$\beta_e$ & 1.3 & 1.4\\
\end{tabular}
\end{ruledtabular}
\end{table}
Simulations A and B2 have been run until the maximum {$J_{rms} = \sqrt{ \langle J^2 \rangle}$ is reached, where $\langle \rangle$ indicates spatial averaging}. This is typically associated with the onset of a fully developed turbulence regime. Runs B1, B3-B6 run for longer, {as can be seen from Table~\ref{tab:sim_datasplit}}.

\subsection{Fluid closure and adiabatic invariants}\label{sec:adiabatic}

Instead of solving the original system of equations~\eqref{eq:vlasov},~\eqref{eq:faraday}, and~\eqref{eq:maxwell}, which is computationally expensive, {it is possible to switch} to fluid quantities.

Defining the number density $n_s(\mathbf{x},t) := \int d^3 v\, f_s(\mathbf{x},\mathbf{v},t)$, we may integrate equation~\eqref{eq:vlasov} over $\int d^3 v$ to obtain the continuity equation
\begin{equation}
    \frac{\partial n_s}{\partial t} + \nabla \cdot  (n_s \mathbf{V}_s) = 0.
\end{equation}
Integrating equation~\eqref{eq:vlasov} over $\int d^3 v \;\mathbf{v}$ and defining $ \mathbf{V}(\mathbf{x},t) := n^{-1}\int d^3 v \; f(\mathbf{x},\mathbf{v},t) \mathbf{v}$ gives the momentum equation
\begin{equation}\label{eq:momentum}
m_s\frac{\partial}{\partial t}\big(n_s\mathbf{V}_s\big) 
+ m_s\,\nabla\!\cdot\!\big(n_s\,\mathbf{V}_s\mathbf{V}_s\big)
= n_s e_s\big(\mathbf{E} + \frac{1}{c} \mathbf{V}_s \times \mathbf{B}\big) 
- \nabla\!\cdot\!\mathbf{P}_s,
\end{equation}
where we have defined the pressure tensor $\textbf{P}_s := {m_s} \int d^3 v \; \left( \mathbf{v}_s - \mathbf{V}_s \right) \left( \mathbf{v}_s - \mathbf{V}_s \right) \, f(\mathbf{x},\mathbf{v}_s,t)$. Neglecting the displacement current in equation~\eqref{eq:maxwell}, defining the bulk (average) plasma velocity as $\mathbf{V}$, and considering the mass ordering $m_e \ll m_i$, we get:
\begin{widetext}
\begin{equation}\label{eq:Ohmslaw}
\mathbf{E} - \frac{d_e^2}{n} \, \Delta \mathbf{E} 
= - \, \mathbf{V} \times \mathbf{B} 
    + \frac{1}{n} \, (\mathbf{J} \times \mathbf{B})
    + \frac{d_e^2}{n} \, \nabla \cdot {\textbf{P}_i} \\[0.3em]
\quad - \frac{1}{n} \, \nabla {\cdot \textbf{P}_e}
    + \frac{d_e^2}{n} \, \nabla \cdot \big[ \mathbf{V} \mathbf{J} + \mathbf{J} \mathbf{V} \big] \\[0.3em]
\quad - \frac{d_e^2}{n} \, \nabla \cdot \left( \frac{\mathbf{J} \mathbf{J}}{n} \right),
\end{equation}
\end{widetext}
which is known as the \emph{generalized Ohm's law}. {Here $\Delta$ stands for Laplacian}. We have written this equation in normalized units, so that spatial coordinates scale as $d_i$, the \emph{ion inertial length}. %, and time units as {$\Omega_{i}$}, the ion cyclotron frequency. 
The terms multiplying ${d^2_e}$, representing the electron inertial length, are related to electron inertia effects. They play an important role in the Electron Diffusion Region (EDR), near the X-point of reconnection events, where they compete with the influence of the electron pressure tensor $\textbf{P}_e$, which is the subject of this manuscript. 

If the generalized Ohm's law, equation~\eqref{eq:Ohmslaw}, is coupled with the fluid or Vlasov equation (equation~\eqref{eq:vlasov}) for ions, under an appropriate choice of $\textbf{P}_e$, the electron pressure tensor, the system is closed. If the ion closure is also prescribed (e.g., at the level of the ion pressure tensor $\textbf{P}_i$), the system is sometimes referred to as eXtended MHD (XMHD) and can be proven to be Hamiltonian~\cite{abdelhamid_hamiltonian_2015,lingam_remarkable_2015}, i.e., possesses structure preservation and even topological invariants~\cite{lingam_concomitant_2016}. 

The standard collisional magnetohydrodynamic condition for $\textbf{P}_e$ corresponds to an adiabatic equation of state, $\textbf{P}_e \sim n^\gamma \textbf{I}$, where $\textbf{I}$ stands for the identity matrix. This relationship breaks down for plasmas that are not in local thermal equilibrium.  {Under the assumption that the magnetic field varies only weakly over a Larmor radius {(an approximation that is reasonable in many smoothly varying magnetized plasmas, but often violated in collisionless reconnection regions)} one may treat the first and second adiabatic invariants as approximately conserved,}
\begin{equation}\label{eq:adiabatic}
\mu := \frac{m V_{\perp}^2}{2 B}=\text { const. } ; \quad \mathcal{J} := m V_{\|} L=\text { const. }
\end{equation}
Indeed, in the absence of Hall, Finite Larmor Radius (FLR) effects and heat flux, one arrives at the well-known CGL (Chew-Goldberger-Low) condition~\cite{chew_boltzmann_1997} $p_{\|} \sim n^3/B^2$ and $p_\perp \sim n B$. See, for instance \citet{hunana_introductory_2019}. 
Below, we consider a more general closure found in seminar works of~\citet{le_equations_2009}
\begin{equation}\label{eq:Le-Egedal}
\begin{aligned}
& \tilde{p}_{\|, e}=\frac{2 {\xi} \alpha}{2 \xi \alpha+1} \frac{\pi \tilde{n}^3}{6 \tilde{B}^2}+\frac{2}{2+\xi \alpha} \tilde{n} ; \\
& \tilde{p}_{\perp, e}=\frac{\xi \alpha}{\xi \alpha+1} \tilde{n} \tilde{B}+\frac{1}{1+\xi \alpha} \tilde{n}
\end{aligned}
\end{equation}

where $\alpha=n_*^3 / B_*^2$, and for any quantity $Q, Q_*=Q / Q_{\infty}$, where $Q_{\infty}$ is the value of $Q$ the reference region away from current sheets. {In this model, the electron equation of state smoothly interpolates between a regime where trapped electrons, with strongly reduced parallel heat conduction, approximately follow CGL‐like double-adiabatic scalings, and a regime where the passing electrons, which retain efficient parallel heat transport along the field, exhibit an almost isothermal (Boltzmann-like) response~\cite{wetherton_validation_2019}. The transition between the regions is} controlled by the parameter $\alpha$. It was originally applied in simpler current sheet conditions. We chose to adapt the model~\eqref{eq:Le-Egedal} by fitting a multiplier $\xi$ in front of the parameter $\alpha$ to better match the data. 

It is possible to obtain more accurate but more expensive 10-moment closures, where the constituent relation is for the heat flux tensor, and thus, pressure is evolved dynamically. Here, we present a version of the heat flux vector under the assumption of a gyrotropic pressure tensor~\cite{hunana_introductory_2019-1}:

\begin{equation}
\begin{aligned}
&\frac{\mathrm{d} p_{\|, \mathrm{e}}}{\mathrm{~d} t}=-p_{\|, \mathrm{e}} \nabla \cdot \boldsymbol{u_{\mathrm{e}}}-2\, p_{\|, \mathrm{e}} \boldsymbol{b} \cdot \nabla \boldsymbol{u_{\mathrm{e}}} \cdot \boldsymbol{b}-\nabla \cdot\left(q_{\|, \mathrm{e}} \boldsymbol{b}\right)+2\, q_{\perp, \mathrm{e}} \nabla \cdot \boldsymbol{b},\\
&\frac{\mathrm{d} p_{\perp, \mathrm{e}}}{\mathrm{~d} t}=-2 p_{\perp, \mathrm{e}} \nabla \cdot \boldsymbol{u}_{\mathrm{e}}+p_{\perp, \mathrm{e}} \boldsymbol{b} \cdot \nabla \boldsymbol{v}_{\mathrm{e}} \cdot \boldsymbol{b}-\nabla \cdot\left(q_{\perp, \mathrm{e}} \boldsymbol{b}\right)-q_{\perp, \mathrm{e}} \nabla \cdot \boldsymbol{b},
\end{aligned}
\end{equation}

In general, at EDR, the electron pressure tensor tends to develop agyrotropy. Following ~\citet{swisdak_quantifying_2016} we define \emph{agyrotropy} as
\begin{equation}\label{eq:agyrotropy}
A=\frac{P_{xy}^2+P_{xz}^2+P_{yz}^2}{P_{\perp}^2+2 P_{\perp} P_{\|}}
\end{equation}
Agyrotropy tends to be present in conjunction with Finite Larmor Radius (FLR) effects.

\begin{comment}
Alternatively, we can construct an empirical formula that is even a better fit using the following anzats 

\begin{equation}
\begin{aligned}\label{eq:CGLplusLIN}
& p_{\|, e}=\alpha \rho_e^\beta B^\gamma+\delta \rho ; \\
& p_{\perp, e}=\varepsilon \rho_e^\zeta B^\eta+\theta \rho_e,
\end{aligned}
\end{equation}
where parameters $\alpha, \beta, \gamma, \delta$ and $\theta$ are all empirically determined. 
\end{comment}

\subsection{Neural closure}\label{sec:neural_closure}
The main goal in this manuscript is to seek a mapping for the electron pressure tensor
\begin{equation}\label{eq:pressure-model}
\mathbf{P} = \mathbf{P}_\theta (n, \mathbf{V}_e, \mathbf{E}, \mathbf{B})
\end{equation}
in terms of lower-order moments and neural network hyperparameters $\theta$. We make no assumption concerning the orientation of the magnetic field to achieve closure, which also works in regions of magnetic field reversals. To achieve such closure, we employ two alternative approaches.

The {Multi-Layer Perceptron (MLP)}~\cite{rosenblatt_perceptron_1958} is an approach identical to that of \citet{laperre_identification_2022}, where the input fields are fed pointwise into the layers of the fully connected neural network (See Figure~\ref{fig:architecture}).

\begin{equation}\label{eq:forwardpass}
\hat{y}_\theta =
f^{(L)} \!\Big(
    W^{(L)} \,
    f^{(L-1)} \!\big(
        \ldots
        f^{(1)} \!\big(
            W^{(1)} \mathbf{x} + \mathbf{b}^{(1)}
        \big)
        \ldots
    \big)
    + \mathbf{b}^{(L)}
\Big)
\end{equation}
where $\mathbf{x}$ are the inputs to the network, $W^{(l)}$ are referred to as weights and $\mathbf{b}$ as biases, while $f$ stands for nonlinear activation functions and $\hat{y}$ is the prediction made by the network. The operation in equation~\eqref{eq:forwardpass} is usually referred to as the \emph{forward pass}.

The Fully Convolutional Neural Network (FCNN) is an architecture consisting solely of convolutional layers (See Figure~\ref{fig:architecture}). Each convolutional layer performs, for each pixel of the image, a multiplication of its value and neighboring values with the corresponding elements of the kernel matrix. Each layer is equipped with a certain number of such kernels that are concatenated together, thus producing a tensor {$W$} whose outer dimensions are referred to as the channel or filter dimensions, whereas the inner indices with respect to which the inner product is taken correspond to spatial dimensions. {Below we provide a mathematical expression for a single convolutional layer with $C_{in}$ channels as inputs, and kernel size $(k_w\times k_h)$}
\begin{equation}
{Y_{o, i, j}=\sum_{c=1}^{C_{i n}} \sum_{u=-k_h/2}^{k_h/2 - 1} \sum_{v=-k_w}^{k_w/2 - 1} W_{o, c, u, v} \times X_{c, i+u, j+v}+b_o}
\end{equation}

The choice of FCNN ensures full translation invariance because only the neighbors of each node are connected. This also makes FCNN much more efficient and lightweight than connecting every point to every other point, which would be impractical and lead to overfitting. To summarize, due to its geometry, FCNN allows us to feed it patches or entire images, rather than points, as in the case of MLP. The way each layer is padded ensures that its output dimension matches its input dimension. In principle, the forward pass can still be formally represented by the operation in equation~\eqref{eq:forwardpass}, but not all weights are allowed. 

The weights and biases are obtained by training the networks on the data, minimizing the Mean Squared Error (MSE) between the predicted values and the actual data (ground truth). 
\begin{equation}\label{eq:MSE}
\text{MSE} = \frac{1}{N} \sum^N_{i=1}\left(y_i - \hat{y}_{i,\theta} \right)^2
\end{equation}
The choice of the loss function is motivated by the fact that MSE corresponds to cross-entropy over a continuous Gaussian variable. We have verified that moments, such as pressure, indeed follow distributions that are close to Gaussian in the data. In principle, more advanced loss functions can be considered in the future to refine the optimization objective. To minimize MSE in equation~\eqref{eq:MSE}, the procedure of \emph{backpropagation} is applied, which is essentially an optimization problem that gradually adjusts the weights of the network according to the gradients of MSE (gradient descent). 
To evaluate the quality of predictions, $R^2$ determination score is computed on validation and test sets:
\begin{equation}
R^2=1-\frac{\sum_i\left(y_i-{\hat{y}_{i,\theta}}\right)^2}{\sum_i\left(y_i-\bar{y}\right)^2},
 \end{equation}
i.e., $R^2$ measures how large MSE is compared to the typical variance in the data~\cite{chicco_coefficient_2021}. Thus, $R^2 = 1$ corresponds to perfect prediction, while $R^2 < 0$ indicates that the error exceeds the typical variance in the data. %Any such prediction from a statistical perspective is considered worse than random guessing. 
We note that here $R^2$ is an ensemble metric that evaluates the quality of predictions pointwise. As such, it is not affected by the image's spatial integrity and is strongly affected by noise in the PIC data, which affects small scales. {We have experimented with other scores such as the Pearson correlation coefficient and the Structural Similarity Index Measure (SSIM), but we found that they often followed the same trends as $R^2$, thus, we decided to keep the study simpler by focusing on one score and visual inspection. }

 \subsection{Datasplit}\label{sec:datasplit}

 \begin{table}[h!]
 \renewcommand{\arraystretch}{1.5}
\centering
\caption{Description of the datasets that are partitioned between test/validation and training. The first column indicates the name of the run, the second column indicates the available and retained timesteps, and the next column indicates which partition this data ends up in. There is also split number 2, which is used solely in the appendix.}\label{tab:sim_datasplit}
\begin{tabular}{l || c c c}
\hline\hline
 & Time steps $\lbrack\omega_{pi}\rbrack$ & \textbf{split 1} & \textbf{split 2} \\
\hline
{PIC A}  &  $375,400,\dots, 750$  & Test 2     & None  \\
PIC B 1 & $375,400,\dots, 1200$ & Test       & Train \\
PIC B 2 & $375,400,\dots, 700\;$ & Validation & Train \\
PIC B 3 & $375,400,\dots,1225$ & Train      & Test \\
PIC B 4 & $375,400,\dots, 1200$ & Train      & Train \\
PIC B 5 & $375,400,\dots, 1050$ & Train      & Validation \\
PIC B 6 & $375,400,\dots,1325$ & Train      & Train \\
\hline
\end{tabular}
\end{table}

 When applying Machine Learning (ML) to physical science, it is crucial to properly split the data into training, validation, and testing sets to prevent data contamination. The training set is the set on which backpropagation and weight and bias optimization are performed. The validation set is the set used to optimize hyperparameters, such as the number of layers and when to stop training. The test set is a separate dataset used to evaluate the performance of several successful ML architectures. There are four levels of data split that can be defined, ranked according to the degree of difficulty. \emph{Random} split involves randomly splitting the data into the aforementioned sets. This approach would be highly problematic for evaluation, as spatially correlated data may be mixed in both training and testing, and the resulting metrics would not be relevant for generalizing to a future state or a new run. \emph{Chronological} split involves taking chunks of temporal data and distributing them accordingly across training, testing, and validation. This is a better approach, but it is most suitable when a stationary regime has been achieved. Nevertheless, there is still no guarantee that satisfactory performance on a chronological split implies that the network will generalize on a new run. \emph{Initialization} split uses an ensemble of runs that share the same governing parameters, and differences arise only from random initialization, turbulence seeds, or noise. This is the approach we take in the current study, where the runs are labeled B1, B2, B3, B4, B5, and B6 (see Table~\ref{tab:sim_datasplit} and the corresponding timeshots used). This way, we can confirm that the methodology generalizes across new initial conditions. %, which is important because the simulation runs for only a few ion cyclotron times, and certain large-scale structures do not have time to change appreciably on those time scales. 

Finally, the most challenging from an ML perspective is the \emph{Out-of-Distribution} (OOD) split. This implies changing the characteristic parameters, such as $\beta_e$ or $\delta B/B$, across validation training/testing (distributing runs with different values of these parameters in training and test sets). From Table~\ref{tab:sim_params} we see that run A has larger values of $\delta B/B$. The difference appears relatively mild, although further analysis reveals that this difference persists over time.

\begin{figure*}[ht!]
\centering
\begin{tabular}{@{}c@{}c@{}c@{}} % The @{} removes inter-column padding

% --- ROW 1 ---
% Each panel is a `subfigure` environment. This is necessary for the \label to work.
% The width is set to a third of the text width to make them fit in a 3-column grid.
\vspace{-.5cm}
\begin{subfigure}{0.33\textwidth}
  % Replace 'example-image-a.png' with your actual PNG file.
  % The [width=\linewidth] option scales the image to the width of the subfigure.
  \begin{overpic}[width=\linewidth]{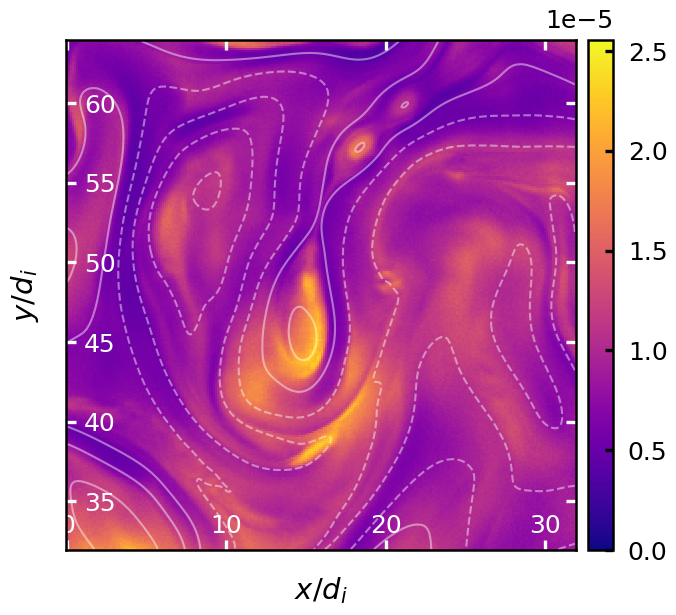}
    % \put(x,y){text}: Places 'text' at coordinates (x,y).
    % The coordinates are percentages of the image width and height.
    % (5,88) means 5% from the left and 88% from the bottom (i.e., top-left).
    % \color{white} makes the text white, which is good for dark backgrounds.
    \put(0,80){\panellabel{a}[black]} 
  \end{overpic}
  \phantomcaption % Empty caption, needed for the subfigure counter and label
  \label{fig:T2D10c1Pxx}
\end{subfigure}&
\begin{subfigure}{0.33\textwidth}
  \begin{overpic}[width=\linewidth]{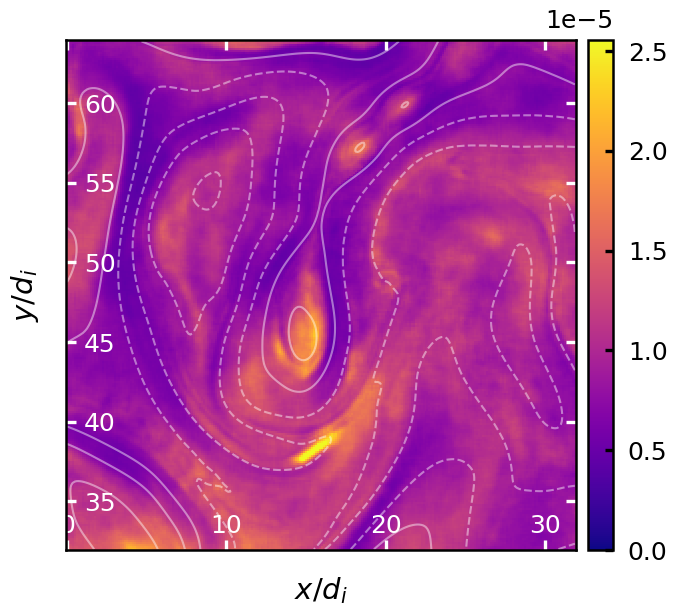}
    \put(0,80){\panellabel{b}[black]} 
  \end{overpic}
  \phantomcaption
  \label{fig:T2D10c1PxxFCNN}
\end{subfigure}&
\begin{subfigure}{0.33\textwidth}
  \begin{overpic}[width=\linewidth]{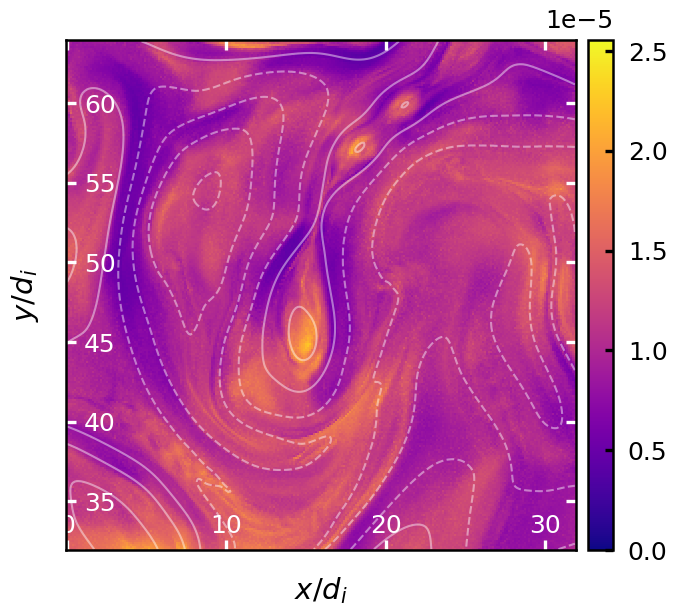}
    \put(0,80){\panellabel{c}[black]} 
  \end{overpic}
  \phantomcaption
  \label{fig:T2D10c1PxxMLP}
\end{subfigure}\\
\vspace{-.25cm}
% --- ROW 2 ---
\begin{subfigure}{0.33\textwidth}
  \begin{overpic}[width=\linewidth]{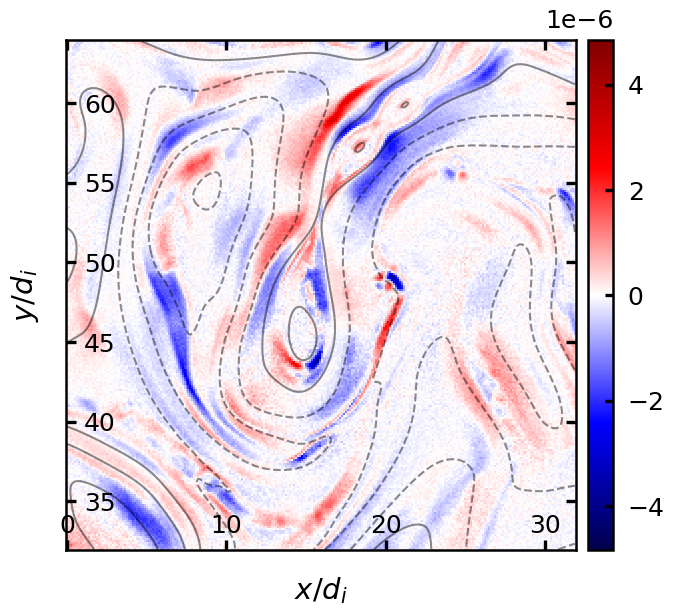}
    \put(0,80){\panellabel{d}[black]} 
  \end{overpic}
  \phantomcaption
  \label{fig:T2D10c1Pxy}
\end{subfigure}&
\begin{subfigure}{0.33\textwidth}
  \begin{overpic}[width=\linewidth]{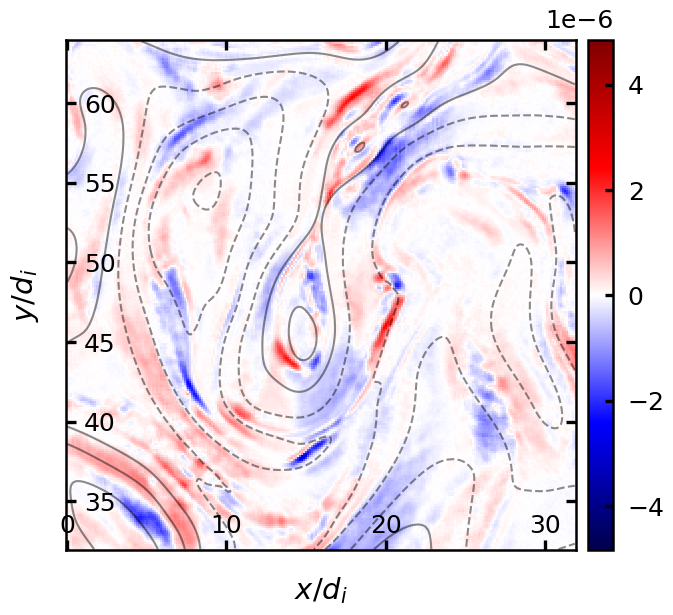}
    \put(0,80){\panellabel{e}[black]} 
  \end{overpic}
  \phantomcaption
  \label{fig:T2D10c1PxyFCNN}
\end{subfigure}&
\begin{subfigure}{0.33\textwidth}
  \begin{overpic}[width=\linewidth]{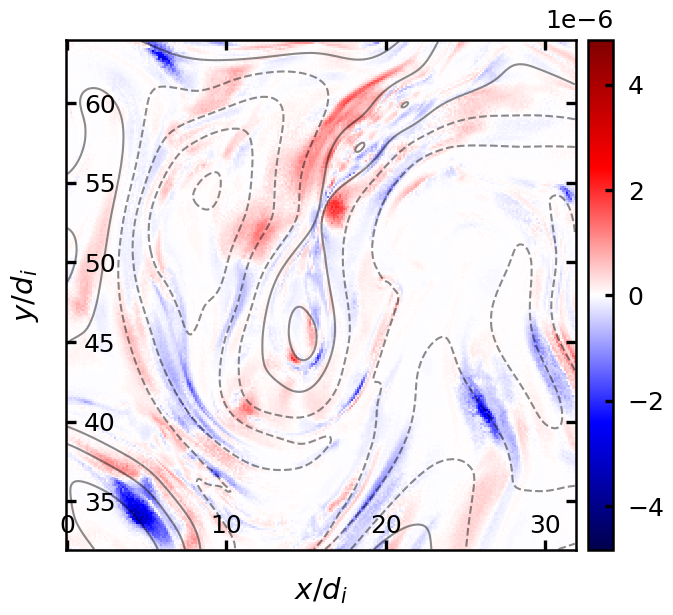}
    \put(0,80){\panellabel{f}[black]} 
  \end{overpic}
  \phantomcaption
  \label{fig:T2D10c1PxyMLP}
\end{subfigure}\\[-1.5ex]

\end{tabular}

\caption{Evaluation of the MLP and FCNN closures on a frame 
$t = 500 \,\omega_{pi}^{-1}$, with run B1 serving as a test set 
(see Table~\ref{tab:sim_datasplit}). 
Rows correspond to different pressure tensor components 
($P_{xx}$, and $P_{xy}$), while columns correspond to 
(a,d) ground truth, (b,e) FCNN predictions, and (c,f) MLP predictions. Each quantity corresponds to the pressure tensor components, with the corresponding color bar on the right. To provide a reference, we add contours of $A_z$, which is equivalent to the flux function in 2D.}
%/volume1/scratch/georgem/closure/models/peppe/sigma0_haydn/FCNN/P/PiD_control.ipynb
\label{fig:pressure_FCNNvsMLP}
\end{figure*}

\subsection{Pressure-strain and scale filtering}\label{sec:pressure_filtering}
Here, we review the Quantities of Interest (QOIs) useful for estimating the transfer of energy between flow, thermal, and electromagnetic {fields} that have been extensively studied in the past by~\citet{yangEnergyTransferChannels2017,yangEnergyTransferPressure2017,matthaeus_pathways_2020}. Using the equations of motion~\eqref{eq:vlasov},~\eqref{eq:faraday} and~\eqref{eq:maxwell}, one can arrive at the following set
\begin{subequations}
\label{eq:energy_channels}
\begin{equation}\label{eq:energy_f}
\partial_t E_s^f+\nabla \cdot\left(E_s^f \boldsymbol{V}_s+\boldsymbol{P}_s \cdot \boldsymbol{V}_s\right)=\left(\boldsymbol{P}_s \cdot \nabla\right) \cdot \boldsymbol{V}_s+\boldsymbol{J}_s \cdot \boldsymbol{E}
\end{equation}
\begin{equation}\label{eq:energy_th}
\partial_t E_s^{th}+\nabla \cdot\left(E_s^{th} \boldsymbol{V}_s+\mathbf{q}_s\right)=-\left(\boldsymbol{P}_s \cdot \nabla\right) \cdot \boldsymbol{V}_s,
\end{equation}
\begin{equation}\label{eq:energy_m}
\partial_t E^m+\frac{c}{4 \pi} \nabla \cdot(\boldsymbol{E} \times \boldsymbol{B})=-\boldsymbol{J} \cdot \boldsymbol{E},
\end{equation}
\end{subequations}
where we follow the standard notation for fluid kinetic energy $E^f_s:= n_s V_s^2/2$, thermal energy $E^{th}_s:= m_s \int d^3 v\,\left(\mathbf{v}-\mathbf{V}_s\right)^2 f_s(\mathbf{x},\mathbf{v},t)/2 $, electromagnetic energy $E^m :=\left(B^2 + E^2\right)/(8\pi)$, current {$J_s := n_s e_sV_s$,} and  heat flux vector $\mathbf{q}_s :=m_s \int d^3 v \, (\mathbf{v}-\mathbf{V}_s)^2 \,(\mathbf{v}-\mathbf{V}_s) \,f_s(\mathbf{x},\mathbf{v},t) /2 $. The terms containing spatial {divergences} on the l.h.s. are responsible for spatial redistribution, whereas the terms of the r.h.s. can be considered as sources and are a major focus. It is convenient to represent the pressure-strain term
\begin{equation}
\begin{aligned}\label{eq:pressure-strain}
-(\boldsymbol{P}_s \cdot \boldsymbol{\nabla}) \cdot \boldsymbol{V}_s & =-p\, \delta_{i j} \,\partial_j V_i^s-\left(P_{i j}^s-p_s \delta_{i j}\right) \,\partial_j V_i^s \\
& =-p_s \,\theta_s-\Pi_{i j}^s\, D_{i j}^s =: -p_s\,\theta_s - PiD_s,
\end{aligned}
\end{equation}
where $p_s:= P_{i i}^s/3$ is the isotropic part, $\quad \Pi_{i j}^s:=P_{i j}^s-p_s \delta_{i j}$ is deviatoric part, $\theta_s:=\nabla \cdot \boldsymbol{V}_s$ is the compressible pressure dilation, and $D_{i j}^s:=(\partial_i V_j^s+\partial_j V_i^s)/2-\theta_s\, \delta_{i j}/3$. The last {term} of equation~\eqref{eq:pressure-strain} is referred to as \emph{pressure-strain} interaction or ``Pi-D''.

{The connection between coherent structures and energy conversion~\cite{yangEnergyTransferChannels2017,yangEnergyTransferPressure2017} can be examined by comparing the spatial distribution of ``Pi-D'' with that of the symmetric velocity stress, the vorticity, and the current density. As scalar measures, we use the normalized quadratic invariants. For the symmetric, traceless rate-of-strain tensor $D_{ij}$ we define}
\begin{equation}\label{eq:Q_D}
Q_D^s=\frac{1}{2} D_{i j}^s D_{i j}^s /\left\langle 2 D_{i j}^s D_{i j}^s\right\rangle,
\end{equation}
{for the voriticity $\boldsymbol{\omega} = \nabla\times \mathbf{v}$,}
\begin{equation}\label{eq:Q_omega}
Q_{\boldsymbol{\omega}}^s=\frac{1}{4} \boldsymbol{\omega}^2_s /\left\langle\boldsymbol{\omega}_s^2\right\rangle,
\end{equation}
{and for the current density}
\begin{equation}\label{eq:Q_j}
Q_J^s=\frac{1}{4} \boldsymbol{J}_s^2 /\left\langle\boldsymbol{J}_s^2\right\rangle.
\end{equation}
{The purpose of this is to perform conditional averages of $PiD$ thresholding above a certain value, which corresponds to identifying coherent structures in the simulation.}
For scale filtering performed on the quantities, see Appendix~\ref{sec:scale2scale}. 

{As mentioned in section~\ref{ref:sim-data} runs B1-B6 have only 256 particles per cell. This generally results in a noisy electric field. To reduce the effects of particle noise, we have applied a box-car filter to smooth all fields across all runs and times using a $4 \times 4$ kernel. This procedure is followed by downsampling from the full-resolution $2048 \times 2048$ to $512 \times 512$.  }

\begin{figure*}[ht!]
\centering
\begin{tabular}{@{}c@{}c@{}c@{}} % The @{} removes inter-column padding

% --- ROW 1 ---
% Each panel is a `subfigure` environment. This is necessary for the \label to work.
% The width is set to a third of the text width to make them fit in a 3-column grid.
%\vspace{-.5cm}
\begin{subfigure}{0.35\textwidth}
  % Replace 'example-image-a.png' with your actual PNG file.
  % The [width=\linewidth] option scales the image to the width of the subfigure.
  \begin{overpic}[width=\linewidth]{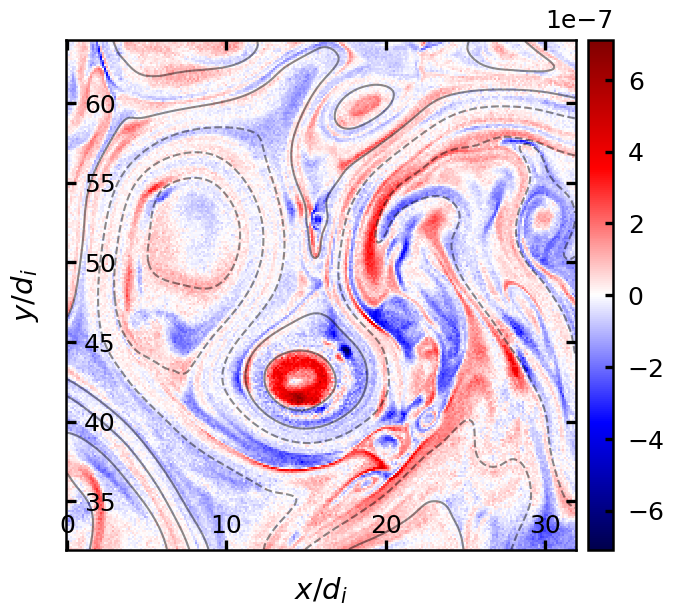}
    % \put(x,y){text}: Places 'text' at coordinates (x,y).
    % The coordinates are percentages of the image width and height.
    % (5,88) means 5% from the left and 88% from the bottom (i.e., top-left).
    % \color{white} makes the text white, which is good for dark backgrounds.
    \put(0,80){\panellabel{a}[black]} 
  \end{overpic}
  \phantomcaption % Empty caption, needed for the subfigure counter and label
  \label{fig:ground_qz}
\end{subfigure}&
\begin{subfigure}{0.35\textwidth}
  \begin{overpic}[width=\linewidth]{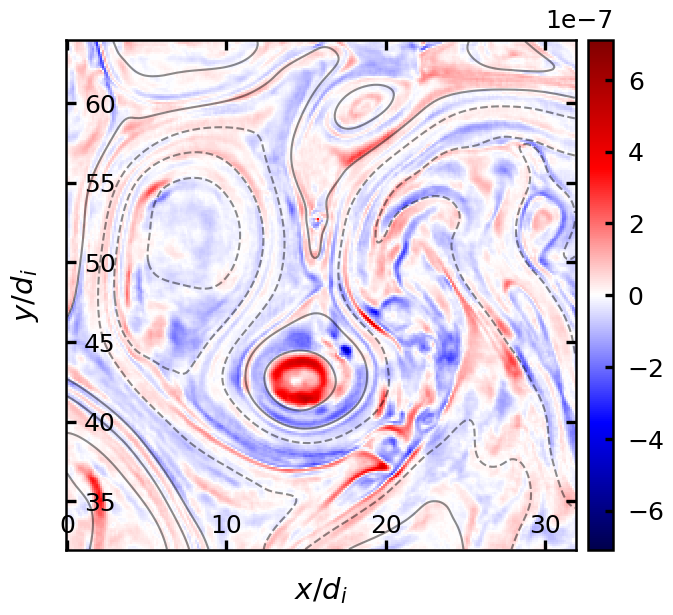}
    \put(0,80){\panellabel{b}[black]} 
  \end{overpic}
  \phantomcaption
  \label{fig:FCNN_qz}
\end{subfigure}
\end{tabular}
%/dodrio/scratch/projects/2024_109/closure/models/peppe/sigma0/FCNN/dist_filter10-16/Q_fromP/4lrs.ipynb
\caption{Plot of z-component of heat flux $q_z$ at $t={825} \; \omega_{pi}^{-1}$ for (a) ground truth, (b) FCNN prediction. Heat flux vector plots are equipped with a corresponding color bar on the right. To provide a reference, we add contours of $A_z$, which is equivalent to the flux function in 2D.}
\label{fig:heat_flux}
\end{figure*}

\section{Results}\label{sec:results}

We obtain pressure and heat flux closure by training the neural networks on data split 1, as shown in Table~\ref{tab:sim_datasplit}. We evaluate the performance by comparing the predicted pressure and derived quantities, such as pressure-strain and agyrotropy, to the \emph{ground truth}, i.e., the actual values in Run B1. In addition, we provide overall statistics of the comparisons, such as the determination score $R^2$. In what follows, we focus on electron closure; thus, when the species index is not specified, we imply $s=e$ for electrons.

\subsection{Pressure and heat flux}\label{sec:pressure_heat}
To provide a baseline model for anisotropic closure, we take the model~\eqref{eq:Le-Egedal} developed by~\citet{le_equations_2009} for parallel $p_\|$ and perpendicular $p_\perp$ pressure, while introducing a fitting parameter $\xi$ to better match the {simulations}. {Since in our domain we have multiple current sheets driven by turbulence rather than a single one~\cite{le_equations_2009}, we choose to normalize asymptotic $\tilde{B}$ and $\tilde{n}$ in equation~\eqref{eq:Le-Egedal} to spatially averaged values. We then fit parallel and perpendicular pressure with a corresponding single value of $\xi$, which is thus a global quantity.}

This model is referred to as ``symbolic'' and is tabulated in Table~\ref{tab:R2_models}, see discussion in section~\ref{sec:adiabatic}. It can be seen that it achieves the same $R^2$ for $p_\perp$ as the simpler double-adiabatic CGL model; however, CGL yields very poor $p_\|$ performance with a negative $R^2$. A negative determination score implies that the model is making predictions that are farther from the ground truth than the typical data variance, indicating that it is not only incorrect conditional on the inputs ($n, B$) but also unconditionally. {Note that here we are referring to a symbolic closure~\cite{le_equations_2009} rather than a machine learning based one, which takes more inputs.}
\begin{table}[h!]
\renewcommand{\arraystretch}{1.5}
\centering
\caption{$R^2$ values of different closure models for parallel, perpendicular, and pressure tensor components. Linear fits of symbolic expressions discussed in section~\ref{sec:adiabatic} with fits performed on the CGL (related to equation~\eqref{eq:adiabatic}), generalized Le-Egedal~\eqref{eq:Le-Egedal}. The fits are performed, showing $R^2$ for parallel and perpendicular pressures.}
\label{tab:R2_models}
% Define column widths: l for the first, and 3 'C' columns of 1.3cm width
\begin{tabular}{|l||C{1.3cm}|C{1.3cm}|C{1.3cm}|C{1.3cm}|C{1.3cm}|}
\hline
\textbf{Model} & CGL & symbolic & MLP & FCNN & B2-trained \\ \hline\hline

% Group 1
$p_\|$ & -0.75 & 0.53 & 0.58 & 0.75 & {0.73} \\
${p_\perp}$ & 0.64 & 0.67 & 0.61 & 0.82    & {0.76} \\ \hline

% Group 2
$P_{xx}$    & --     & --   & 0.59 & 0.82      & {0.74} \\
$P_{yy}$   & --      & --   & 0.57 & 0.82 & {0.74}  \\
$P_{zz}$    & --     & --   & 0.57 & 0.80        & {0.72}  \\ \hline

% Group 3
$P_{xy}$   & --      & --   & -0.04 & 0.37  & {0.18}  \\
$P_{xz}$   & --      & --   & 0.01  & 0.41  & {0.17}\\
$P_{yz}$   & --      & --   & 0.00  & 0.39  & {0.19} \\ \hline

% Group 4 
$q_{x}$   & --      & --   & {0.03} & 0.20 & --\\
$q_{y}$   & --      & --   & {0.04}  & 0.23 & --\\
${q_z}$   & --      & --   & {0.06}  & 0.30 & --\\ \hline
%/dodrio/scratch/projects/2024_109/closure/models/peppe/sigma0/FCNN/dist_filter10-16/Q_fromP/4lrs.ipynb
\end{tabular}
\end{table}

In Table~\ref{tab:R2_models}, we also compare the results of the symbolic fit to those of the Multi-Layer Perceptron (MLP) and Fully Convolutional Neural Network (FCNN) introduced in section~\ref{sec:neural_closure} with the pipeline graphically represented in Figure~\ref{fig:architecture}. From Table~\ref{tab:R2_models} we see that evaluation of MLP and symbolic model~\eqref{eq:Le-Egedal} results in comparable $p_\|$ and $p_\perp$. In contrast, FCNN outperforms MLP on all metrics, yielding relatively good $R^2 \gtrsim 0.8$ for diagonal components of the pressure tensor. FCNN yields below average $R^2 \sim 0.4$ determination score for off-diagonal components, which significantly outperforms MLP $R^2 \sim 0$. 

To provide spatial characteristics of the neural closures, in Figure~\ref{fig:pressure_FCNNvsMLP}, we plot pressure components $P_{xx}$ and $P_{xy}$ from a subset of the simulation at $t=500\omega_{pi}^{-1}$, focusing on an island chain that has just undergone reconnection. Fully developed turbulence is typically identified by the maximum of the Root Mean Square (RMS) current, which occurs at $t = 550\omega_{pi}^{-1}$, indicating we are near this regime. We see that FCNN prediction for $P_{xx}$ on Figure~\ref{fig:T2D10c1PxxFCNN} matches Figure~\ref{fig:T2D10c1Pxx} better than MLP prediction on Figure~\ref{fig:T2D10c1PxxMLP}. For instance, a ridge just outside of the main magnetic island at (15 $d_i$,37 $d_i$) is missed by the MLP. In addition, FCNN is more faithful to the overall structures, but we note small-scale vapor-like artifacts that are not present in the original $P_{xx}$ or the MLP prediction. In contrast, MLP prediction appears smoother. The scale of the artifacts is below $d_i$ {and it does not consistently reach the scale of the grid}. 

FCNN prediction on Figure~\ref{fig:T2D10c1PxyFCNN} for $P_{xy}$ captures overall features of the ground truth on Figure~\ref{fig:T2D10c1Pxy} significantly better than MLP on Figure~\ref{fig:T2D10c1PxyMLP}. For instance, the positive $P_{xy}$ at the separatrix at ($16 d_i, 53 d_i$) is more accurately reproduced with FCNN, while MLP tends to predict sign reversal incorrectly. FCNN is more accurate at reproducing the signs of the ridges around the magnetic island ($15 d_i,37 d_i$). Nevertheless, one cannot help but notice certain small-scale irregularities in FCNN prediction, which can also be described as vapor-like noise. These features are consistent with overall scores reported in the Table~\ref{tab:R2_models}. 

We complete this subsection by reporting on the spatial structures of FCNN predicted heat flux $q_z$ in Figure~\ref{fig:FCNN_qz}. Despite the relatively low $R^2$ score reported in the Table~\ref{tab:R2_models}, we see that FCNN captures main structures, for instance, counter streams of $q_z$ inside magnetic islands, and even structures around the ridges. Upon closer inspection, we see the same vapor-like noise, which likely contributes to a low $R^2$ score. {In Table~\ref{tab:R2_models} we have also included comparison of overall $R^2$ obtained with FCNN and with MLP for $\textbf{q}$. We see that MLP is much inferior at reconstructing the heat flux closure.}

\subsection{Characteristics of anisotropies}\label{sec:anisotropies}

\begin{figure*}[ht!]
\centering
\begin{tabular}{@{}c@{}c@{}c@{}} % The @{} removes inter-column padding

% --- ROW 1 ---
% Each panel is a `subfigure` environment. This is necessary for the \label to work.
% The width is set to a third of the text width to make them fit in a 3-column grid.
\vspace{-.5cm}
\begin{subfigure}{0.33\textwidth}
  % Replace 'example-image-a.png' with your actual PNG file.
  % The [width=\linewidth] option scales the image to the width of the subfigure.
  \begin{overpic}[width=\linewidth]{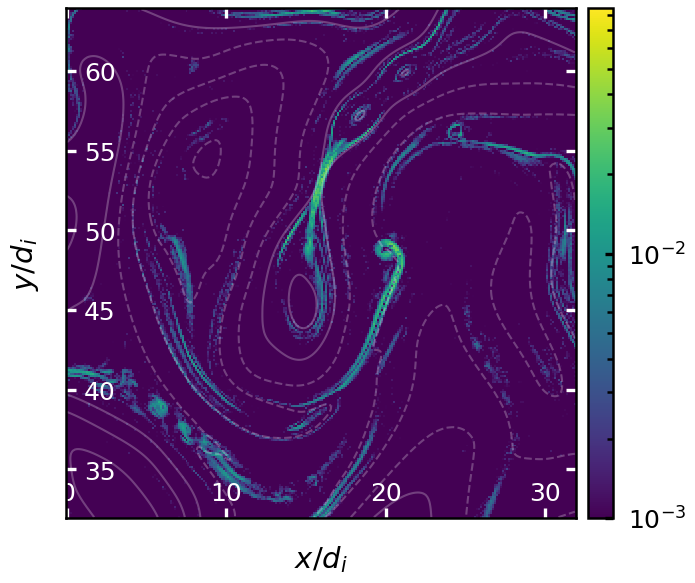}
    % \put(x,y){text}: Places 'text' at coordinates (x,y).
    % The coordinates are percentages of the image width and height.
    % (5,88) means 5% from the left and 88% from the bottom (i.e., top-left).
    % \color{white} makes the text white, which is good for dark backgrounds.
    \put(0,80){\panellabel{a}[black]} 
  \end{overpic}
  \phantomcaption % Empty caption, needed for the subfigure counter and label
  \label{fig:ground_agyrotropy}
\end{subfigure}&
\begin{subfigure}{0.33\textwidth}
  \begin{overpic}[width=\linewidth]{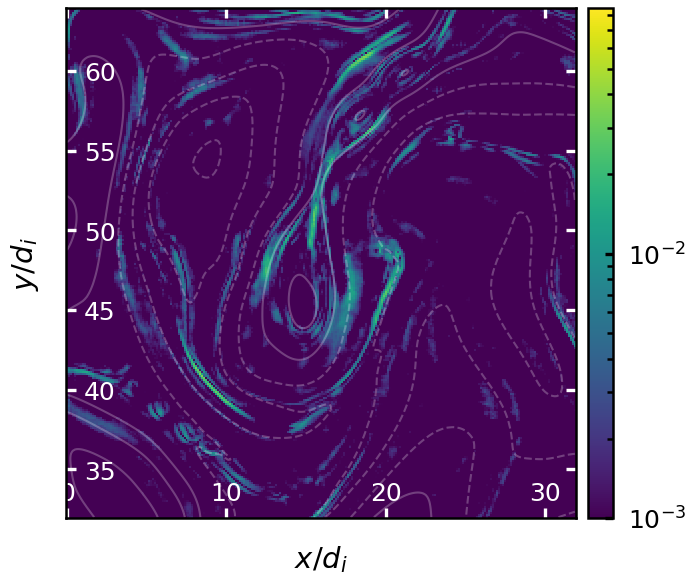}
    \put(0,80){\panellabel{b}[black]} 
  \end{overpic}
  \phantomcaption
  \label{fig:FCNN_agyrotropy}
\end{subfigure}&
\begin{subfigure}{0.33\textwidth}
  \begin{overpic}[width=\linewidth]{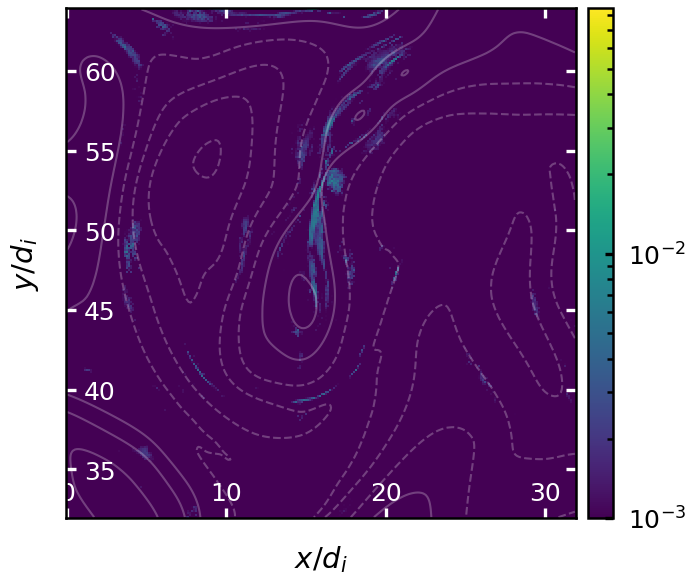}
    \put(0,80){\panellabel{c}[black]} 
  \end{overpic}
  \phantomcaption
  \label{fig:MLP_agyrotropy}
\end{subfigure}\\

% --- ROW 3 ---
\begin{subfigure}{0.33\textwidth}
  \begin{overpic}[width=\linewidth]{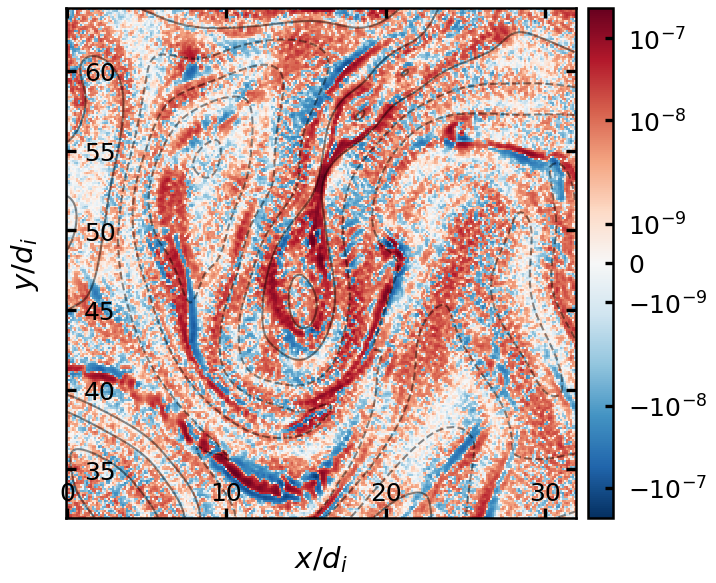}
    \put(0,80){\panellabel{d}[black]} 
  \end{overpic}
  \phantomcaption
  \label{fig:MLPc1Pxx}
\end{subfigure}&
\begin{subfigure}{0.33\textwidth}
  \begin{overpic}[width=\linewidth]{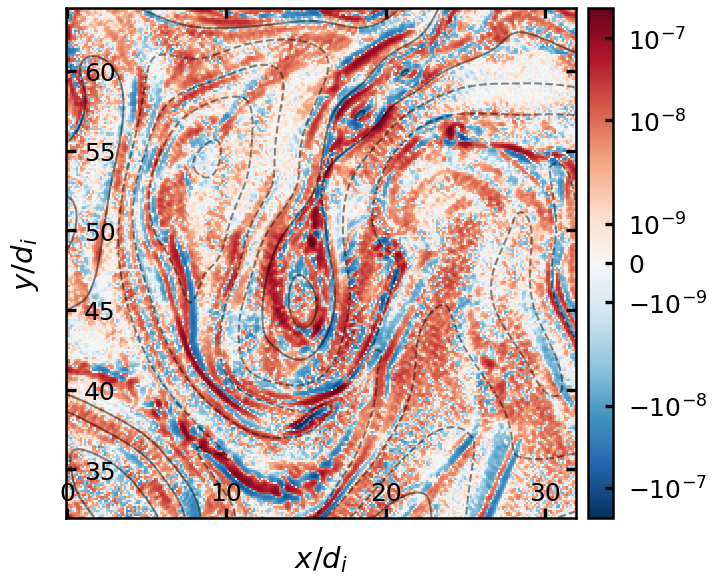}
    \put(0,80){\panellabel{e}[black]} 
  \end{overpic}
  \phantomcaption
  \label{fig:MLPc1Pxy}
\end{subfigure}&
\begin{subfigure}{0.33\textwidth}
  \begin{overpic}[width=\linewidth]{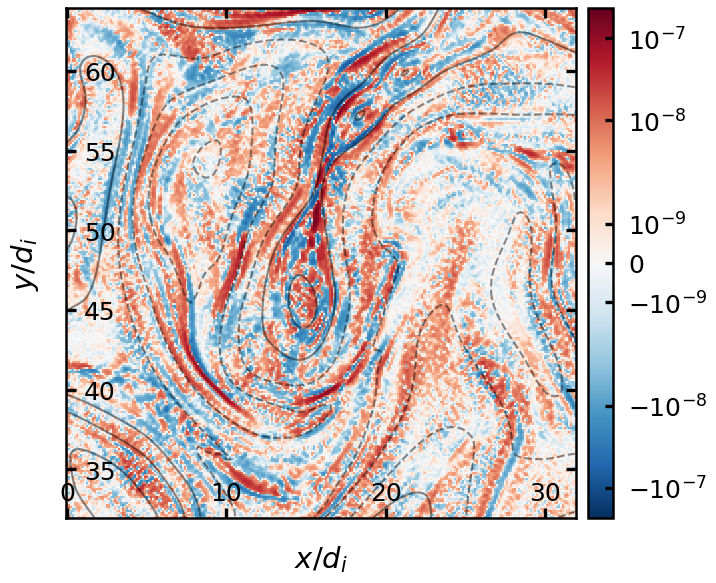}
    \put(0,80){\panellabel{f}[black]} 
  \end{overpic}
  \phantomcaption
  \label{fig:MLPc1PiD}
\end{subfigure}
%\vspace{-.25cm}

\end{tabular}
%/volume1/scratch/georgem/closure/models/peppe/sigma0/MLP/dist_filter10-16/P/PiD_control.ipynb
\caption{Plot of agyrotropy (equation~\eqref{eq:agyrotropy}) and incompressible pressure-strain $PiD$ at $t = 500 \; \omega_{pi}^{-1}$ for (a) agyrotropy ground truth, (b) FCNN prediction of agyrotropy, (c) MLP prediction of agyrotropy, (d) $PiD$ ground truth, (e) FCNN prediction of $PiD$, (f) MLP prediction of $PiD$. To provide a reference, we add contours of $A_z$, which is equivalent to the flux function in 2D.}
\label{fig:agyrotropy}
\end{figure*}

We utilize the synthetic electron pressure predicted by FCNN and MLP, and apply it {along} with {the predictor variables such as} lower-order moments $n_e, \mathbf{V}_e, \,\mathbf{E},\, \mathbf{B}$ to compute derived quantities.
We plot ground truth agyrotropy defined in equation~\eqref{eq:agyrotropy} in Figure~\ref{fig:ground_agyrotropy} at $t=500 {\omega_{pi}^{-1}}$ at the same location as Figure~\ref{fig:pressure_FCNNvsMLP}. We observe that agyrotropy is strongest at the X point and along the separatrices. It also takes large values near the ridge surrounding the principal magnetic island in the island chain. This general structure is replicated in the FCNN closure estimated agyrotropy on Figure~\ref{fig:FCNN_agyrotropy}. However, MLP closure hardly predicts any agyrotropy in Figure~\ref{fig:MLP_agyrotropy}, which is consistent with $R^2\sim 0$ for off-diagonal elements as indicated in Table~\ref{tab:R2_models}.

\begin{figure*}[ht!]
\centering
\begin{tabular}{@{}c@{}c@{}c@{}} % The @{} removes inter-column padding

% --- ROW 1 ---
% Each panel is a `subfigure` environment. This is necessary for the \label to work.
% The width is set to a third of the text width to make them fit in a 3-column grid.
\vspace{-.5cm}
\begin{subfigure}{0.33\textwidth}
  % Replace 'example-image-a.png' with your actual PNG file.
  % The [width=\linewidth] option scales the image to the width of the subfigure.
  \begin{overpic}[width=\linewidth]{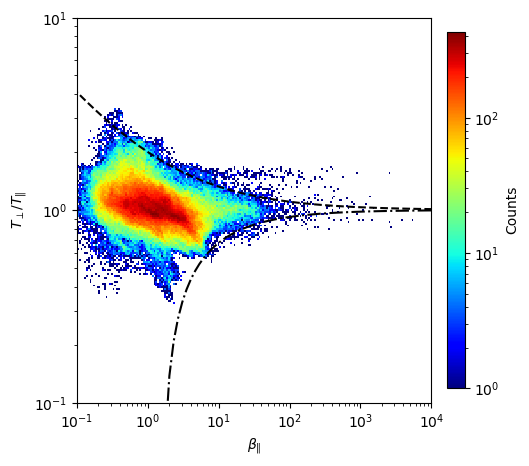}
    % \put(x,y){text}: Places 'text' at coordinates (x,y).
    % The coordinates are percentages of the image width and height.
    % (5,88) means 5% from the left and 88% from the bottom (i.e., top-left).
    % \color{white} makes the text white, which is good for dark backgrounds.
    \put(0,80){\panellabel{a}[black]} 
  \end{overpic}
  \phantomcaption % Empty caption, needed for the subfigure counter and label
  \label{fig:brazil_ground}
\end{subfigure}&
\begin{subfigure}{0.33\textwidth}
  \begin{overpic}[width=\linewidth]{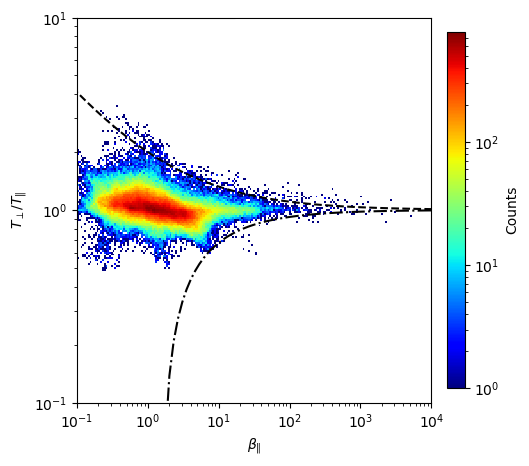}
    \put(0,80){\panellabel{b}[black]} 
  \end{overpic}
  \phantomcaption
  \label{fig:brazil_FCNN}
\end{subfigure}&
\begin{subfigure}{0.33\textwidth}
  \begin{overpic}[width=\linewidth]
{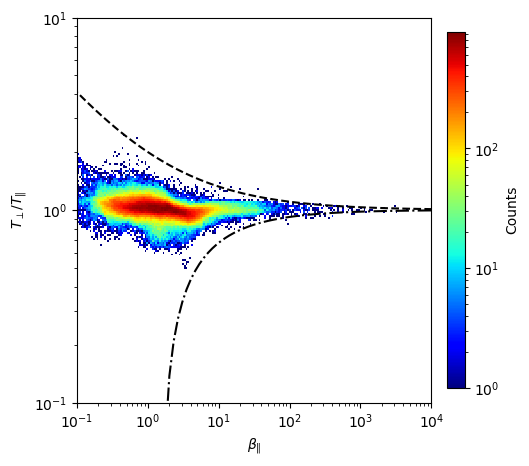}
    \put(0,80){\panellabel{c}[black]} 
  \end{overpic}
  \phantomcaption
  \label{fig:brazil_MLP}
\end{subfigure}\\
\vspace{-.25cm}

\end{tabular}
%/volume1/scratch/georgem/closure/models/peppe/sigma0/MLP/dist_filter10-16/P/PiD_control.ipynb
\caption{Temperature anisotropy vs. $\beta_\|$ plots (a.k.a. Brazil plot) histograms with counts represented on the rainbow colormap for (a) ground truth.  (b) FCNN prediction. (c) MLP prediction. The dashed line corresponds to the onset of whistler instability, while the dot-dashed line corresponds to the onset of the firehose instability. {The functional form of the curves is given by the expression $T_\perp/T_\| = 1 + A/\beta_{e,\|}^B$, where we have taken $A = 1, B = 0.49$ for whistlers~\cite{gary_whistler_1996} (dashed line) and $A = 1.32, B = 0.61$ for electron firehose~\cite{gary_resonant_2003} (dot-dashed line). These parameters correspond to specific growth rates, $\gamma = 0.1 \,{\Omega_{e}}$}}
\label{fig:brazil}
\end{figure*}

As is well known, anisotropies lead to microinstabilities, such as the whistler instability~\cite{gary_linear_2006}, which constrain {anisotropies}. This is usually illustrated by plotting~\cite{hellinger_solar_2006} the anisotropy $T_\|/T_\perp$ versus $\beta_\|$ as we do on Figure~\ref{fig:brazil}. Figure~\ref{fig:brazil_ground} corresponding to ground truth shows that {the anisotropy parameter space} is constrained from above by whistler instability and from below by electron firehose instability, with some traces outside those ranges. Concurrently, FCNN on Figure~\ref{fig:brazil_FCNN} predicts a similar distribution respecting the instability thresholds, but it appears to reduce the variance of $T_\|/T_\perp$ anisotropy somewhat. The pressure computed by MLP has an even smaller range, as shown in Figure~\ref{fig:brazil_MLP}, which bolsters previous results comparing the two architectures. {The fact that we observe lower values of anisotropies for FCNN and especially MLP is most likely due to the fact that there is regression to the mean, i.e., the neural network has a tendency to predict fewer extreme values since it was trained to optimize mean square error. } 

\subsection{Energy channels}\label{sec:channel_evaluations}
\begin{figure*}[ht!]
\centering
\begin{tabular}{@{}c@{}c@{}c@{}} % The @{} removes inter-column padding

% --- ROW 1 ---
% Each panel is a `subfigure` environment. This is necessary for the \label to work.
% The width is set to a third of the text width to make them fit in a 3-column grid.
\raisebox{-.3cm}
{
\begin{subfigure}{0.33\textwidth}
  % Replace 'example-image-a.png' with your actual PNG file.
  % The [width=\linewidth] option scales the image to the width of the subfigure.
  \begin{overpic}[width=\linewidth]{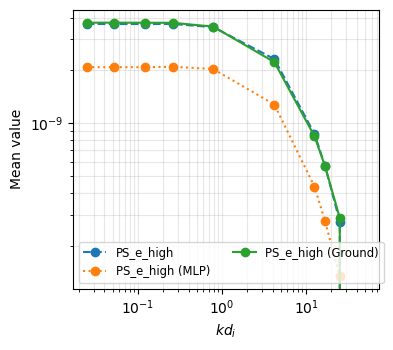}
  %/Users/u0167590/github/closure/examples/scale2scale.ipynb
    % \put(x,y){text}: Places 'text' at coordinates (x,y).
    % The coordinates are percentages of the image width and height.
    % (5,88) means 5% from the left and 88% from the bottom (i.e., top-left).
    % \color{white} makes the text white, which is good for dark backgrounds.
    \put(5,80){\panellabel{a}[black]} 
  \end{overpic}
  \phantomcaption % Empty caption, needed for the subfigure counter and label
  \label{fig:PShighpass}
\end{subfigure}}&
\begin{subfigure}{0.32\textwidth}
  \begin{overpic}[width=\linewidth]{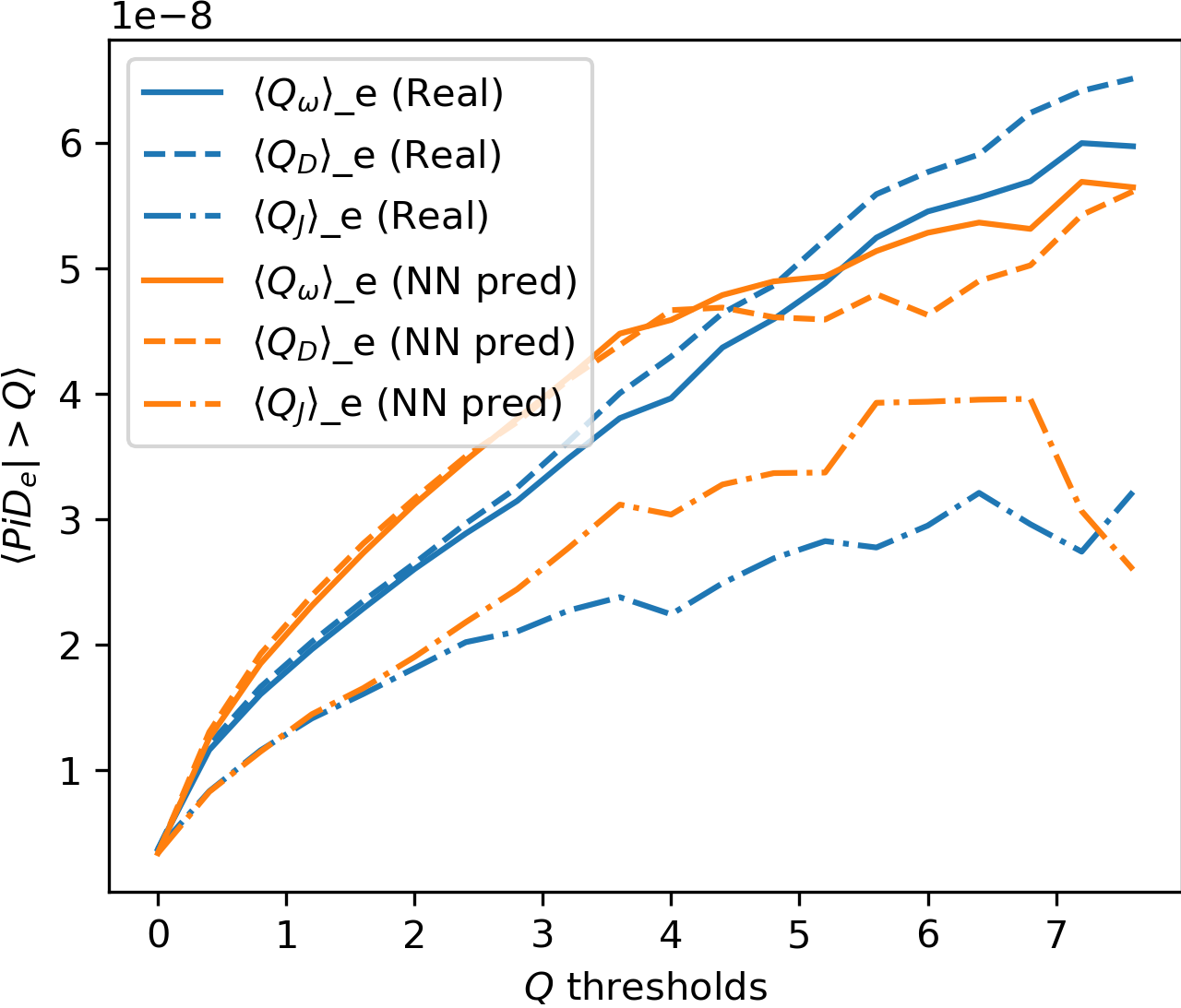}
    \put(-1,80){\panellabel{b}[black]} 
  \end{overpic}
  \phantomcaption
  \label{fig:FCNN_PiD_threshold}
\end{subfigure}&
\begin{subfigure}{0.32\textwidth}
  \begin{overpic}[width=\linewidth]{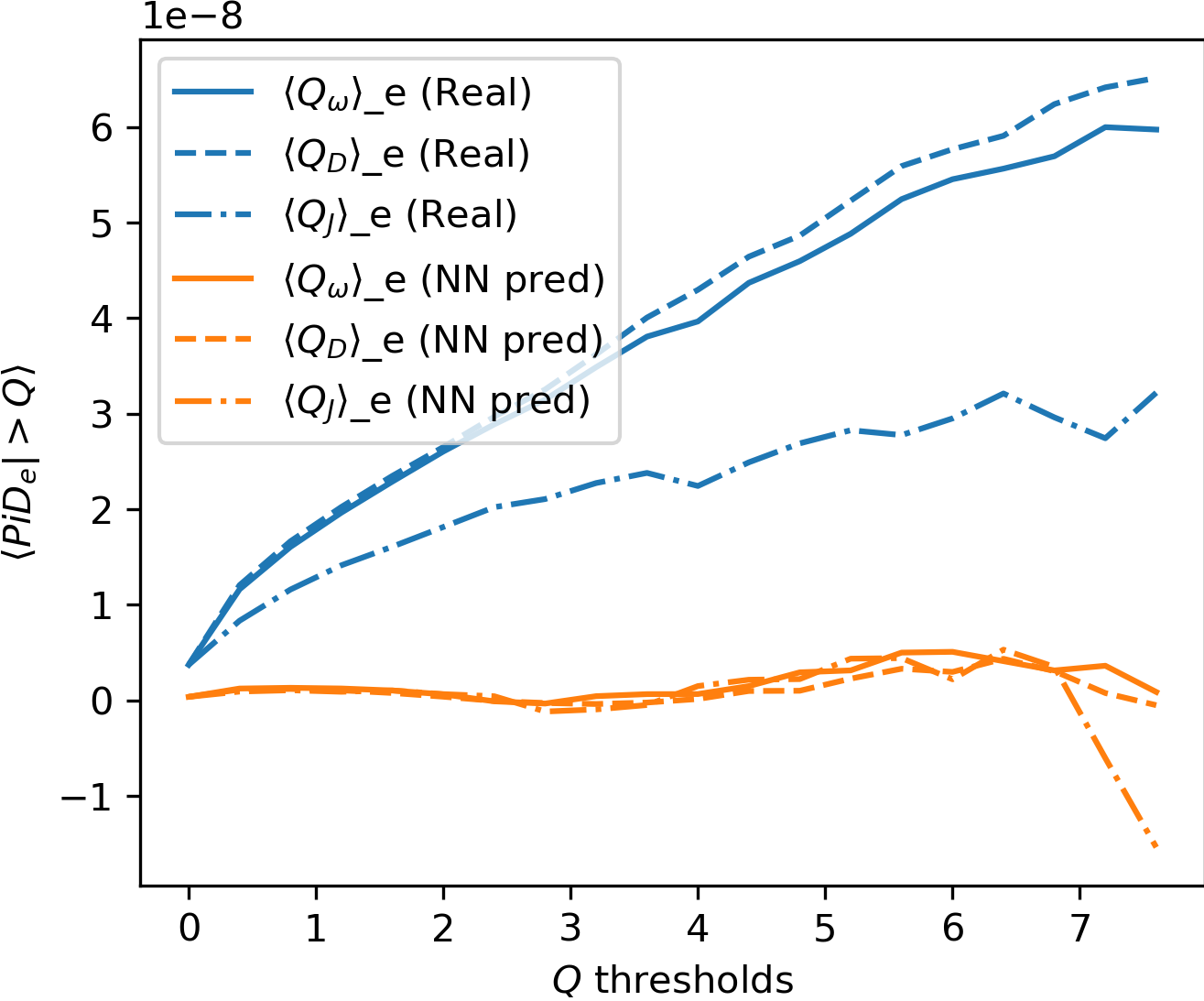}
    \put(3,80){\panellabel{c}[black]} 
  \end{overpic}
  \phantomcaption
  \label{fig:MLP_PiD_threshold}
\end{subfigure}\\
%/volume1/scratch/georgem/closure/models/peppe/sigma0_haydn/FCNN/P/PiD_control.ipynb
%/Users/u0167590/github/closure/examples/scale2scale.ipynb
\end{tabular}
%/volume1/scratch/georgem/closure/models/peppe/sigma0/MLP/dist_filter10-16/P/PiD_control.ipynb
\caption{(a) Total pressure-strain $PS$ with high pass filter $k^\star$ wavenumber applied. On the x-axis, we plot $k^\star$. The blue dashed curve shows the FCNN prediction, the green solid curve shows the ground truth, and the orange dotted curve shows the MLP prediction. (b) and (c): Average incompressible pressure-strain $PiD$, conditional to values of quantities $Q_\omega$, $Q_D$ and $Q_j$ above a certain threshold (see equations~\eqref{eq:pressure-strain},~\eqref{eq:Q_D}) - dashed line,~\eqref{eq:Q_omega} - solid line and~\eqref{eq:Q_j} - dot dashed line. This quantity is computed for the ground truth (blue curves) and neural network prediction (orange curves). Note that in panel b, the orange curve corresponds to the predictions made with the help of FCNN, while in panel c, it corresponds to the predictions made with the help of MLP. }
\label{fig:cond_pres_strain}
\end{figure*}

Energy channels discussed in section~\ref{sec:pressure_filtering} express the conversion of flow~\eqref{eq:energy_f} into thermal energy~\eqref{eq:energy_th} signified by the pressure-strain term as well as more conventional Ohmic dissipation that converts electromagnetic energy~\eqref{eq:energy_m} into flow~\eqref{eq:energy_f}. First, we are going
to investigate scale-to-scale k-filtering defined in Appendix~\ref{sec:scale2scale} and applied to pressure-strain $PS(k > k_c) $~\eqref{eq:pressure-strain}, {i.e., with} a high-pass filter at scale $k_c$. We plot $PS(k > k_c) $ on Figure~\ref{fig:PShighpass}, which demonstrates that overall, both FCNN and MLP capture the general distribution of pressure-strain over the scales, with FCNN shadowing accurately the quantity and MLP underestimating it by a factor of 2. 

Since pressure-strain quantifies the particle heating, it is of interest to investigate whether it peaks on coherent structures, as was done in~\citet{yangEnergyTransferChannels2017,yangEnergyTransferPressure2017}. For this aim, we compute the average incompressible portion of pressure-strain $\langle PiD | Q > Q^\star \rangle $, where $Q$ stands for any of the three quantities in equation~\eqref{eq:Q_D},~\eqref{eq:Q_omega},~\eqref{eq:Q_j}, each represented by dashed, dot-dashed, and solid lines on Figure~\ref{fig:FCNN_PiD_threshold}, which compares ground truth to FCNN and Figure~\ref{fig:MLP_PiD_threshold} for MLP. We note the similarity of $\langle PiD | Q > Q^\star \rangle$, which tends to be larger for larger thresholds, consistent with \citet{yangEnergyTransferChannels2017,yangEnergyTransferPressure2017}. However, in this case, the same trend is also observed for $Q_J$ conditionals, indicating the association of current sheets and heating in our simulations. The comparison reveals qualitatively similar behavior for FCNN closure but rather poor results for MLP closure, which completely underestimates conditionals of $\langle PiD | Q > Q^\star \rangle$. 

\subsection{Ablation study and generalization}\label{sec:ablation}

Our aim here is two-fold. First, we would like to perform an \emph{ablation study}, i.e., remove certain features (inputs such as $n_e, \mathbf{V}_e, \mathbf{E}, \mathbf{B}$ from $\mathbf{P}_e = \mathbf{P}_{NN}(n_e, \mathbf{V}_e, \mathbf{E}, \mathbf{B})$ and {train from scratch} the FCNN. This gives as a collection of models that we refer to as  \textit{default} := $\mathbf{P}_{NN}(n_e,\mathbf{V}_e,\mathbf{E},\mathbf{B})$, \textit{noE} := $\mathbf{P}_{NN}(n_e,\mathbf{V}_e,\mathbf{B})$, \textit{Jtot} := $\mathbf{P}_{NN}(n_e,\mathbf{J},\mathbf{E},\mathbf{B})$, \textit{JtotnoE} := $\mathbf{P}_{NN}(n_e,\mathbf{J},\mathbf{B})$ and allows us to study the influence of each feature in predicting our target $\mathbf{P}$. Secondly, we would like to study the generalization of the neural network trained, validated, and tested on B1-B6 to run {A} (see Table~\ref{tab:sim_datasplit}). To this end, we plot the results of the ablation/generalization study in Figure~\ref{fig:ablation_study_heatmap}. It is organized into three panels: Figure~\ref{fig:ablation_T2D10c1} applies the test set B1, Figure~\ref{fig:ablation_data} applies the test set {A}, and Figure~\ref{fig:ablation_comparison} illustrates the difference between the two. 

The conclusion from Figure~\ref{fig:ablation_T2D10c1} is that test set B1 is less sensitive towards the collection of ablation models. The worst-performing (marginally) model is \textit{noE}, for instance the \textit{default} results in $R^2 = 0.82$ for $p_\perp$, while \textit{noE} yields $0.73$. Likewise \textit{default} model yields $R^2 = 0.75$ for $p_\|$, while \textit{noE} gives $0.68$. The effect is strongest for diagonal components. This indicates that the {e}lectric field contains useful information for predicting the heating of plasma, albeit marginally. On the other hand, when the same set of models is applied to test set {A}, a different picture is observed in Figure~\ref{fig:ablation_data}. The most noticeable is the complete reversal of the performance of the \textit{noE} model, which turns out to be the most performant model for the diagonal part of the pressure tensor. For instance, it yields $R^2 = 0.75$ for $p_\perp$, while \textit{default} only results in $R^2 = 0.47$. Similarly, \textit{default} yields $R^2 = 0.61$ for $p_\|$, while \textit{noE} yields $R^2 = 0.61$. The worst-performing model is \textit{Jtot} with $R^2=0.15$ for $p_\|$ and $R^2 = 0.25$ for $p_\perp$. The off-diagonal components of the pressure tensor are largely unaffected. The conclusion that can also be drawn from Figure~\ref{fig:ablation_comparison} is that \textit{noE} is the most statistically robust model; therefore, we use it for the remainder of the manuscript. {We provide the following explanation for why $noE$ model performs better. Run A has a much less noisy electric field than runs B1-B6. This means there is a significant shift in the distribution of $\mathbf E$ when comparing A and B. The currently used 4$\times $4 box filter was insufficient to address this problem; thus, we propose dropping $\mathbf E$ from the predictors.
}

Next, we investigate the fidelity of the spatial structures of the FCNN closure of \textit{noE} model in Figure~\ref{fig:generalization}. The ground truth that comes from run {A} consists of several current sheets that have already become unstable. The choice of the time snapshot also occurs near the maximum $J_{rms}$, like in Figure~\ref{fig:pressure_FCNNvsMLP}. The most interesting one is located at $(6 d_i, 35 d_i)$ and leads to enhancement of heating as can be inferred from large values of $P_{xx}$ in Figure~\ref{fig:data_ground_Pxx}. In Figure~\ref{fig:data_ground_Pxy}, strong positive and negative values of $P_{xy}$ are seen at the right separatrix, and a negative enhancement of $P_{xy}$ is seen just north of the X point. The inspection of Figure~\ref{fig:data_FCNN_Pxy} at that location reveals a similar pattern of $P_{xy}$, albeit less intense. In general, $P_{xy}$ served by FCNN appears less intense and a bit more patchy, but even some small-scale structures coincide. We turn towards comparison between the incompressible part of pressure-strain $PiD$ on Figure~\ref{fig:data_ground_PiD} and the predicted on Figure~\ref{fig:data_FCNN_PiD}. We see that in both cases, $PiD$ tends to have large positive values on the separatrices. Comparison between other structures is also consistent, with some exceptions. For instance, the X point at $15 d_i, 60 d_i$ has a mismatch in the polarity of $PiD$.

{We also study the Galilean invariance of the \emph{noE} model in appendix~\ref{ref:galilean}

\begin{figure*}[ht!]
\centering
\begin{tabular}{@{}c@{}c@{}c@{}} % The @{} removes inter-column padding

% --- ROW 1 ---
% Each panel is a `subfigure` environment. This is necessary for the \label to work.
% The width is set to a third of the text width to make them fit in a 3-column grid.
\vspace{-.5cm}
\begin{subfigure}{0.33\textwidth}
  % Replace 'example-image-a.png' with your actual PNG file.
  % The [width=\linewidth] option scales the image to the width of the subfigure.
  \begin{overpic}[width=\linewidth]{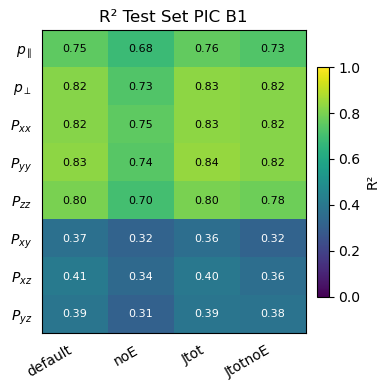}
  %/Users/u0167590/github/closure/examples/scale2scale.ipynb
    % \put(x,y){text}: Places 'text' at coordinates (x,y).
    % The coordinates are percentages of the image width and height.
    % (5,88) means 5% from the left and 88% from the bottom (i.e., top-left).
    % \color{white} makes the text white, which is good for dark backgrounds.
    \put(0,95){\panellabel{a}[black]} 
  \end{overpic}
  \phantomcaption % Empty caption, needed for the subfigure counter and label
  \label{fig:ablation_T2D10c1}
\end{subfigure}&
\begin{subfigure}{0.33\textwidth}
  \begin{overpic}[width=\linewidth]{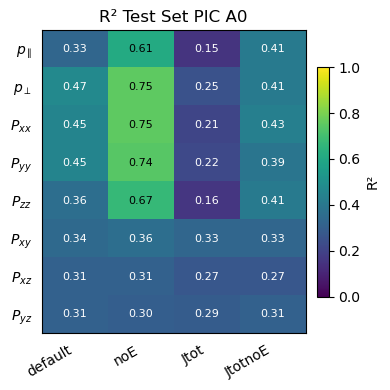}
    \put(0,95){\panellabel{b}[black]} 
  \end{overpic}
  \phantomcaption
  \label{fig:ablation_data}
\end{subfigure}&
\begin{subfigure}{0.33\textwidth}
  \begin{overpic}[width=\linewidth]{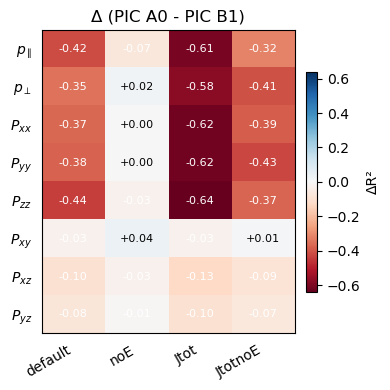}
    \put(0,95){\panellabel{c}[black]} 
  \end{overpic}
  \phantomcaption
  \label{fig:ablation_comparison}
\end{subfigure}\\
\end{tabular}
%/volume1/scratch/georgem/closure/models/peppe/sigma0/MLP/dist_filter10-16/P/PiD_control.ipynb
\caption{Ablation study of input configurations for the FCNN. Panels show $R^{2}$ score tables for (a) Test Set PIC B1, (b) Test Set PIC {A}, and (c) their difference $\Delta= \text{{A}}-\text{B1}$ for each predicted quantity (rows) and model variant (columns). Each panel consists of a grid with rows labelled by specific quantity that is predicted, such as $p_\|,p_\perp,\dots$ while the columns are labelled by different models which are distinguished by their inputs: \textit{default} = $(n_e,\mathbf{V}_e,\mathbf{E},\mathbf{B})$, \textit{noE} = $(n_e,\mathbf{V}_e,\mathbf{B})$, \textit{Jtot} = $(n_e,\mathbf{J},\mathbf{E},\mathbf{B})$, \textit{JtotnoE} = $(n_e,\mathbf{J},\mathbf{B})$. Warmer colors indicate higher $R^{2}$, see the color bar; in the difference panel on the right, red (blue) denotes improvement (degradation) on PIC B1 relative to PIC {A}. }
 \label{fig:ablation_study_heatmap}
    %/Users/u0167590/github/closure/examples/scale2scale.ipynb
    %/volume1/scratch/georgem/closure/models/peppe/sigma0_haydn/FCNN/PnoE/PiD_control.ipynb
    %/volume1/scratch/georgem/closure/models/peppe/sigma0_haydn/FCNN/PnoE/NN2data.ipynb
    %/volume1/scratch/georgem/closure/models/peppe/sigma0_haydn/FCNN/PJtotnoE/PiD_control.ipynb
    %/volume1/scratch/georgem/closure/models/peppe/sigma0_haydn/FCNN/PJtotnoE/NN2data.ipynb
    %/volume1/scratch/georgem/closure/models/peppe/sigma0_haydn/FCNN/PJtot/PiD_control.ipynb
    %/volume1/scratch/georgem/closure/models/peppe/sigma0_haydn/FCNN/PJtot/NN2data.ipynb
    %/volume1/scratch/georgem/closure/models/peppe/sigma0_haydn/FCNN/P/NN2data.ipynb
    %/volume1/scratch/georgem/closure/models/peppe/sigma0_haydn/FCNN/P/PiD_control.ipynb
    %/volume1/scratch/georgem/closure/models/peppe/sigma0_haydn/FCNN/PJtotnoE/NN2data.ipynb
\end{figure*}

We conclude the results section with the analog of Figure~\ref{fig:FCNN_PiD_threshold} with the same analysis performed to obtain $\langle PiD | Q > Q^\star \rangle$, applied to the dataset {A} and plotted on Figure~\ref{fig:data_PiD_thresholds}. The blue lines represent the ground truth, showing consistency with Figure~\ref{fig:FCNN_PiD_threshold} in regards to $Q_\omega$ and $Q_D$ conditionals, but showing that the {A} dataset has stronger $PiD$ conditionals to $Q_J$ and diverges even more from the result of \citet{yangEnergyTransferChannels2017,yangEnergyTransferPressure2017}. The prediction of FCNN matches the $\langle PiD | Q > Q^\star \rangle$ behavior qualitatively but shows a different large $Q$ threshold tail for all three conditionals. We note, however, that for large $Q > 4$ this statistic is not reliable because the larger the $Q$, the fewer pixels are actually used to compute these quantities of interest.

\begin{figure*}[ht!]
\centering
\begin{tabular}{@{}c@{}c@{}c@{}} % The @{} removes inter-column padding

% --- ROW 1 ---
% Each panel is a `subfigure` environment. This is necessary for the \label to work.
% The width is set to a third of the text width to make them fit in a 3-column grid.
%\vspace{-.5cm}
\begin{subfigure}{0.33\textwidth}
  % Replace 'example-image-a.png' with your actual PNG file.
  % The [width=\linewidth] option scales the image to the width of the subfigure.
  \begin{overpic}[height=.8\textwidth]{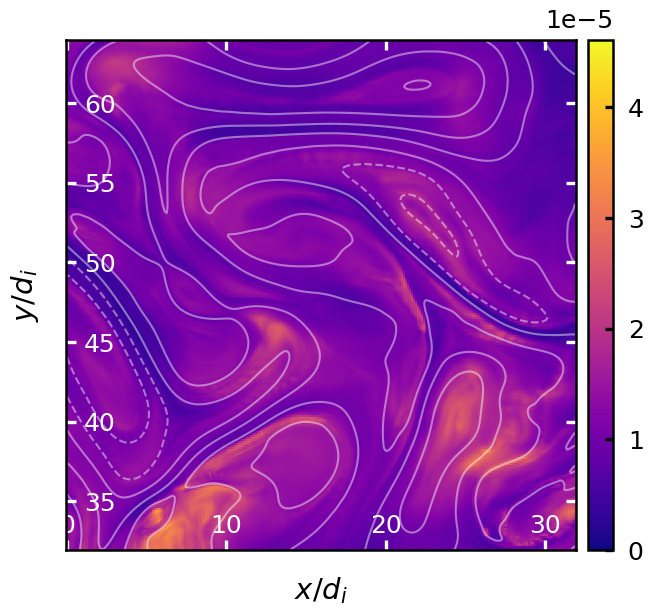}
    % \put(x,y){text}: Places 'text' at coordinates (x,y).
    % The coordinates are percentages of the image width and height.
    % (5,88) means 5% from the left and 88% from the bottom (i.e., top-left).
    % \color{white} makes the text white, which is good for dark backgrounds.
    \put(0,80){\panellabel{a}[black]} 
  \end{overpic}
  \phantomcaption % Empty caption, needed for the subfigure counter and label
  \label{fig:data_ground_Pxx}
\end{subfigure}&
\begin{subfigure}{0.33\textwidth}
  \begin{overpic}[height=.8\textwidth]{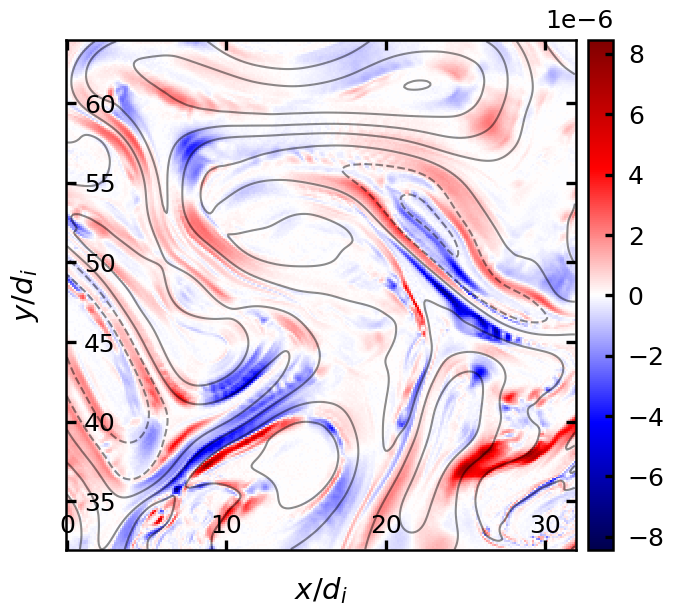}
    \put(0,80){\panellabel{b}[black]} 
  \end{overpic}
  \phantomcaption
  \label{fig:data_ground_Pxy}
\end{subfigure}&
%\raisebox{-.5cm}
{%
\begin{subfigure}{0.33\textwidth}
  \begin{overpic}[height=.8\textwidth]{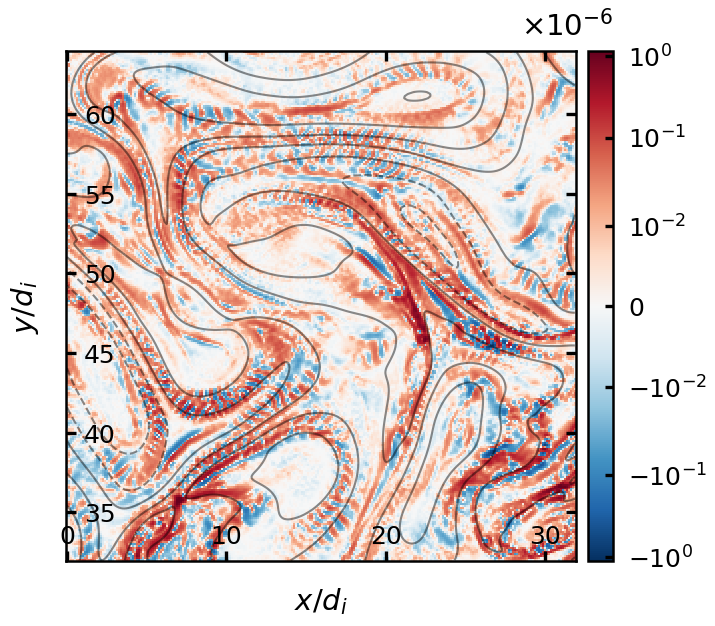}
    \put(0,75){\panellabel{c}[black]} 
  \end{overpic}
  \phantomcaption
  \label{fig:data_ground_PiD}
\end{subfigure}}\\
%\vspace{-.25cm}
% --- ROW 2 ---
\begin{subfigure}{0.33\textwidth}
  \begin{overpic}[height=.8\textwidth]{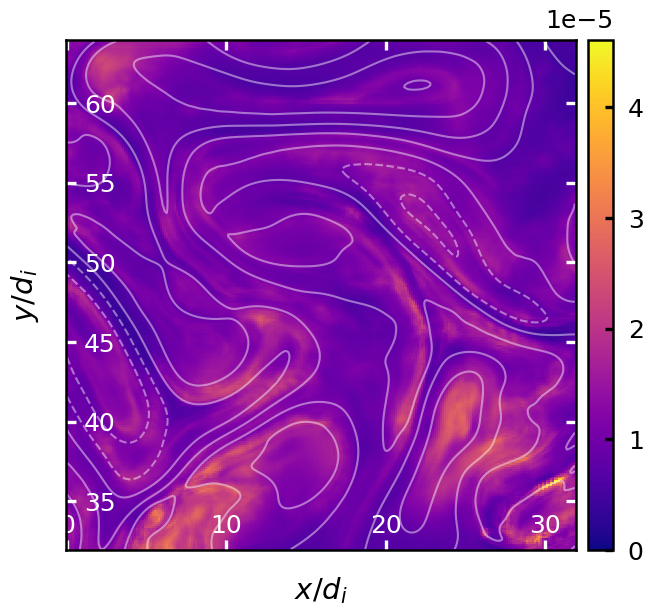}
    \put(0,80){\panellabel{d}[black]} 
  \end{overpic}
  \phantomcaption
  \label{fig:data_FCNN_Pxx}
\end{subfigure}&
\begin{subfigure}{0.33\textwidth}
  \begin{overpic}[height=.8\textwidth]{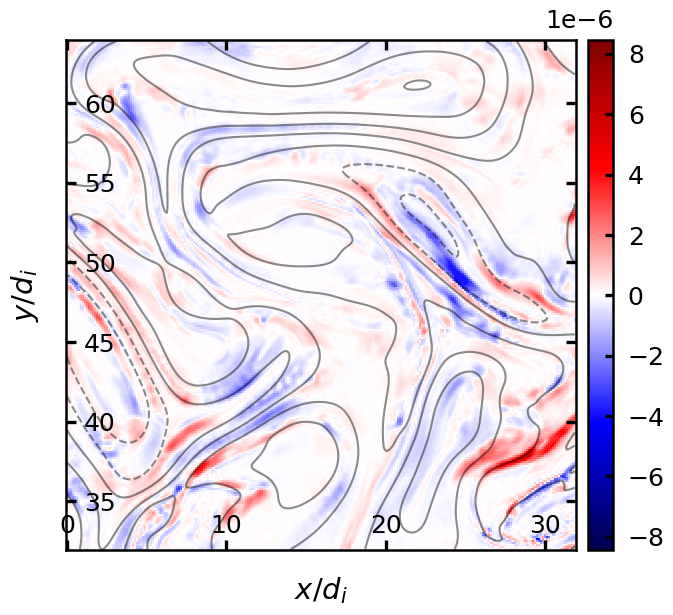}
    \put(0,80){\panellabel{e}[black]} 
  \end{overpic}
  \phantomcaption
  \label{fig:data_FCNN_Pxy}
\end{subfigure}&
%\raisebox{-.5cm}
{\begin{subfigure}{0.33\textwidth}
  \begin{overpic}[height=.8\textwidth]{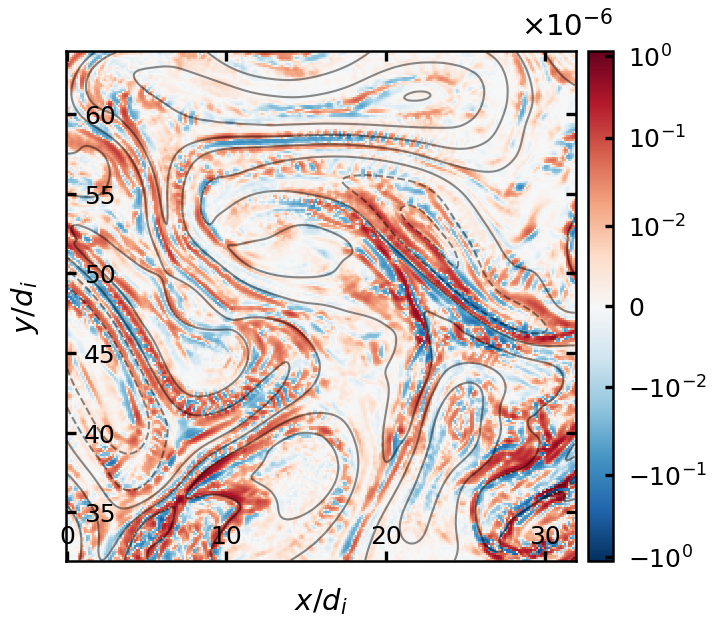}
    \put(0,75){\panellabel{f}[black]} 
  \end{overpic}
  \phantomcaption
  \label{fig:data_FCNN_PiD}
\end{subfigure}}%\\[-1.5ex]
\end{tabular}
%/volume1/scratch/georgem/closure/models/peppe/sigma0_haydn/FCNN/PnoE/NN2data.ipynb
\caption{Evaluation of the FCNN closures on a frame $t=525 \,\omega_{pi}^{-1}$, with run A serving as a test set (see Table~\ref{tab:sim_datasplit}).  (a) Ground truth $P_{xx}$. (b) Ground truth $P_{xy}$. (c) Ground truth $PiD$. (d) FCNN predicted $P_{xx}$. (e) FCNN predicted $P_{xy}$. (f) FCNN predicted $PiD$. Each panel is equipped with the corresponding colormap. To provide a reference, we add contours of $A_z$, which is equivalent to the flux function in 2D.}
\label{fig:generalization}
\end{figure*}

\begin{figure}
    \centering
    \includegraphics[width=0.9\linewidth]{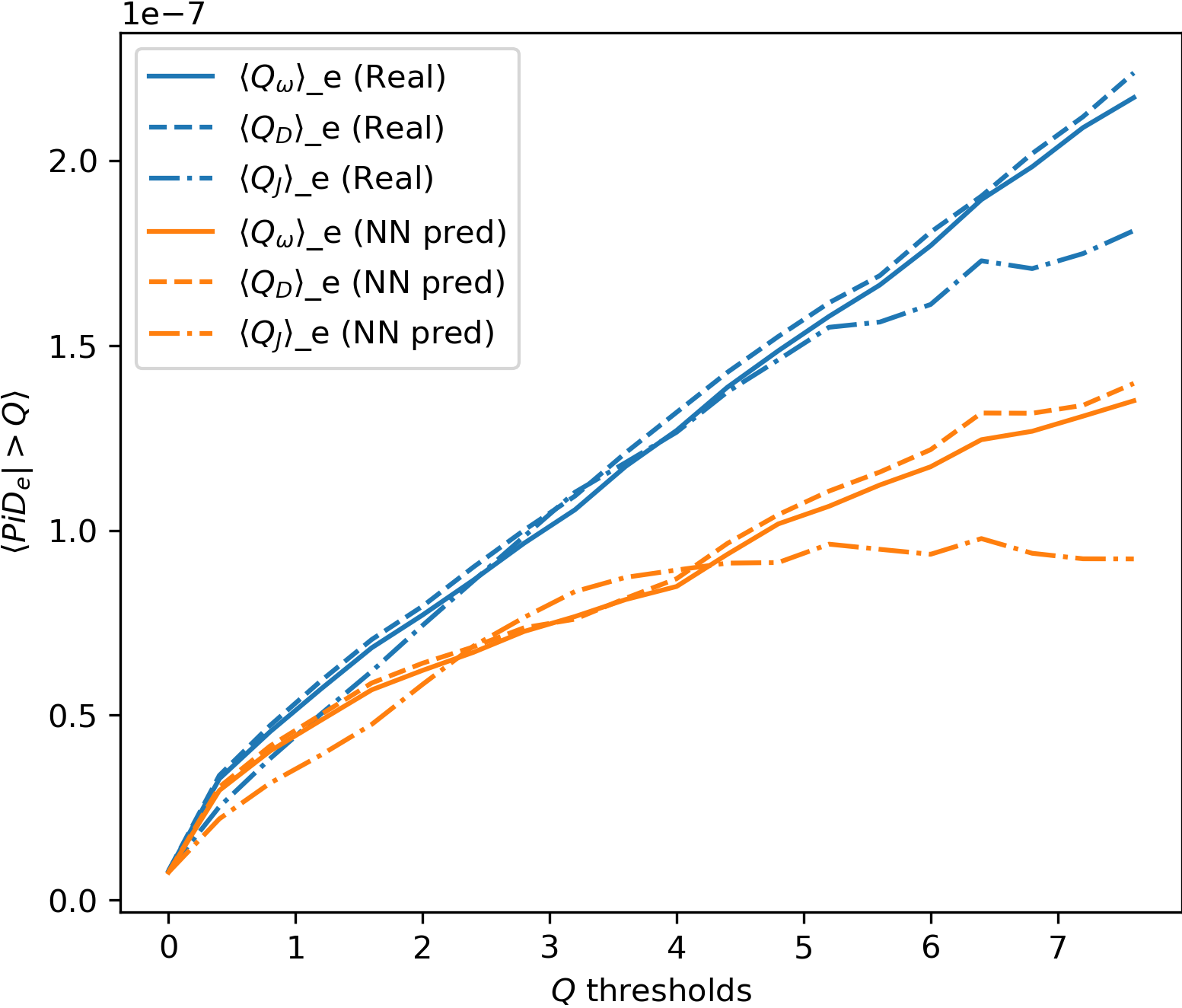}
    \caption{Average incompressible pressure-strain $PiD$, conditional to values of quantities $Q_\omega$, $Q_D$ and $Q_J$ above a certain threshold (see equations~\eqref{eq:pressure-strain},~\eqref{eq:Q_D}) - dashed line,~\eqref{eq:Q_omega} - solid line and~\eqref{eq:Q_j} - dot dashed line. This quantity is computed for the ground truth (blue curves) and neural network prediction (orange curves).}
    \label{fig:data_PiD_thresholds}
\end{figure}

%%%%%%%%%%%%%%%%%%%%%%%%%%%%%%%%%%%%%%%

%%%%%%%%%%%%%%%%%%%%%%%%%%%%%%%%%%%%%%%

\section{Discussion}\label{sec:discussion}

In this manuscript, we introduce a new non-local neural closure for the electrons in the turbulent magnetosheath. This closure is obtained by training a Fully Convolutional Neural Network (FCNN) on the output of a fully kinetic Particle-in-Cell simulation of the energy-conserving code ECsim~\cite{lapenta_exactly_2017}. We submit this closure to multiple statistical evaluations on a hold-out test set, which comes from two separate numerical simulations. One test set comes from run B1 (tabulated in Table~\ref{tab:sim_datasplit}) with the exact same ECsim parameters but a different random initialization. The other test set comes from run {A}, which is a simulation performed earlier~\cite{arro_spectral_2022} with a much larger number of particles and slightly different physical parameters (see Table~\ref{tab:sim_params}). Because of this performance on run {A} can be considered a generalization test.

The closure we obtain is a five-moment closure for the electron pressure tensor as a function of density, electron velocity, magnetic field, and electric field. It can be considered a neural generalization of trapped/passing particle closure $p = p(n, B)$~\cite{le_equations_2009} (see equation~\eqref{eq:Le-Egedal}), which is a symbolic expression and is itself a generalization of CGL~\cite{chew_boltzmann_1997}. We also compare it to architecture similar to the one used by \citet{laperre_identification_2022}, which consists of a Multi-Layer Perceptron (MLP), a more traditional neural network that we apply locally, pointwise. In contrast, FCNN is applied to patches and can thus be considered a non-local closure. We benchmark the performance of FCNN, MLP, and more traditional symbolic anisotropic $p_\|$ and $p_\perp$ closures~\cite{le_equations_2009}. We also attempted to learn a 10-moment closure for the heat flux using an FCNN architecture, but we found the performance to be relatively poor, which could be attributed to the limited data available. In general, we found that the 5-moment pressure tensor closure performed significantly worse when trained on a single simulation than when trained on four simulations. 

Generally speaking, for the diagonal elements, we find that MLP provides results identical to symbolic closure~\cite{le_equations_2009}, but FCNN identifies many of the mesoscale structures more accurately (see section~\ref{sec:pressure_heat}), while introducing some vapor-like noise. Furthermore, MLP does not capture well the quantities related to off-diagonal elements of the pressure tensor, such as agyrotropy (see section~\ref{sec:anisotropies}). MLP underestimates the distributional spread in anisotropies, as well as derived quantities such as pressure-strain (section~\ref{sec:channel_evaluations}), which is quite important for estimating the energy budget and energy channels in turbulence. In contrast, FCNN closure captures the important structures in the off-diagonal pressure tensor, albeit not perfectly. This is confirmed by lower values of determination score $R^2\sim 0.4$ compared to diagonal components $R^2\sim 0.75$. Nevertheless, the derived quantities from the FCNN-computed pressure tensor, such as the spatial distribution of pressure-strain and scale-to-scale filtered incompressible pressure-strain, appear quite similar to the results from Direct Numerical Simulation (DNS). In particular, the budget of mean incompressible pressure-strain PiD, conditional on coherent structures defined according to the approach of~\citet{yangEnergyTransferChannels2017,yangEnergyTransferPressure2017}, reveals agreement between the actual and predicted values by the FCNN. The mean scale-to-scale filtered distribution of pressure-strain over the spatial scales, down to the smallest scales, is also consistent when comparing FCNN to the ground truth, while MLP tends to underestimate it.

To gain insight into the importance of pressure-strain features, such as the electric field, for instance, we train a selected set of pressure tensor models with different inputs in section~\ref{sec:ablation}. Of particular interest are the following models: \emph{default}, which is the standard model that takes as input all the lower-order moments, and \emph{noE}, which omits the electric field. These models are evaluated on two sets, {A} and B1, corresponding to very large and small number of particles per cell, respectively (see Table~\ref{tab:sim_params}). We find that \emph{default} drastically under-performs on the generalization data set {A}, while the score for \emph{noE} is essentially unchanged. This is confirmed when carefully inspecting the spatial structures of $P_{xx}$, $P_{xy}$, and $PiD$ associated with current sheets in the simulation. This leads to the conclusion that the electric field is not a reliable predictor of the closure, at least when the closure is trained on more noisy simulations in the set B. {Note that as mentioned in section~\ref{sec:pressure_filtering}, we attempted to reduce this noise by applying Gaussian filters. For the discussion on generalization with respect to Galilean boosts, see Appendix~\ref{ref:galilean}.}

The fact that \emph{noE} is able to generalize to the simulation with a different number of particles is reassuring, since certain small-scale structures are not consistent across the runs, and indicates that when the input parameters are chosen properly, the closure appears robust. We emphasise that this was achieved for only one set of physical parameters, namely $\delta B/B \sim 0.6-0.7$, $\beta_i \sim 5.3-5.7$, and $\beta_e \sim 1.3-1.4$, consistent with magnetosheath conditions, and in 2D. In the future, we plan to train on a broader set of conditions, consisting of a parameter sweep over these quantities, to find a closure that interpolates between these regimes and extend our method to 3D. Generally, such parameter changes entail \emph{out-of-distribution shifts} (OOD) in the relevant quantities such as density or pressure. This problem is usually treated with \emph{transfer learning}. This implies taking a pre-trained base network and fine-tuning some of the network's layers in response to distribution shifts in the new physical conditions. This approach has shown promise in Large Eddy Simulations (LES) in the work of~\citet{subel_explaining_2023} across several cases: adapting to changes in the forcing wavelength and increasing the Reynolds number. We propose undertaking such work in plasma physics in the future. The problem of introducing a neural network into a parametrization of a physical process is a challenging one and requires a series of adaptations, including training on multiple-step roll-outs~\cite{frezat_posteriori_2022,subel_explaining_2023}. This implies wrapping a numerical solver inside the loss function of the optimization algorithm, a method also referred to as \emph{online (a posteriori) training and testing}. {Recently, there have been developments on promoting globally stable data-driven models, which can be achieved via modifications of the objective function~\cite{kaptanoglu_promoting_2021} or via enforcing hyperbolicity~\cite{christlieb_hyperbolic_2025}}

Additional steps can be taken in the future to improve the quality of the pressure tensor closure. It is clear that with more training data, better results may be achieved; however, this implies computational costs associated with generating such data, especially since some runs must be reserved for validation. In principle, many groups worldwide possess Particle-in-Cell or Vlasov simulation data that could be useful for training such closures. Thus, one could conceive of deeper neural network architectures trained on the wealth of simulation data, mirroring the work in meteorology where neural surrogate models were trained on 40 years of reanalysis data, rivaling numerical weather prediction models~\cite{kochkov_neural_2024,pathak_fourcastnet_2022}. {The latter is}, of course, possible thanks to the existence of high-quality data obtained on assimilating observations into models~\cite{hersbach_era5_2020}, a dataset of such quality we do not possess in the context of space plasmas. Thus, from this angle, developing data-driven models that outperform existing high-fidelity numerical simulations such as ECsim~\cite{markidis_energy_2011,lapenta_exactly_2017} does not appear to be feasible in the near future; however, training neural surrogates that are more efficient than conventional methods can still be achieved. In particular, the closures obtained in this way can be embedded in Reduced Order Models (ROMs) such as fluid~\cite{lani_gpu-enabled_2014} and hybrid kinetic models~\cite{behar_menura_2022,palmroth_vlasov_2018} that are more computationally tractable than fully kinetic simulations and with the appropriate closure can reproduce the outputs of such fully kinetic simulations~\cite{wang_comparison_2015,ohia_demonstration_2012,ng_simulations_2017,finelli_bridging_2021}. Similar closures allow the study of kinetic processes such as Kinetic Alfen Wave (KAW) turbulence on a larger span of inertial range~\cite{miloshevich_modeling_2019,miloshevich_inverse_2021}. The goal of data-driven closure is to extend the validity of ROMs, which, thanks to their efficiency, can be simulated more frequently and across a broader range of parameters than high-fidelity models.

We would like to emphasize a few other avenues that could guide data-driven closure development. First, in this manuscript, we have experimented with rather traditional architectures, such as MLP and FCNN, which leave room for more modern AI models, including neural operators, such as Fourier Neural Operators (FNOs) ~\cite{anandkumar_neural_2020,li_fourier_2021}. FNO has already been applied for neural surrogate modelling of a plasma fusion device~\cite{gopakumar_fourier_2023}. Furthermore, we have relied solely on the standard loss function, Mean Squared Error (MSE), and have not exploited \emph{soft constraints}~\cite{beucler_enforcing_2021}, additional physics-based constraints that can enhance the physical fidelity of the learned representation. %Embedding equations in losses has become very popular after the emergence of Physics Informed Neural Networks (PINNs) ~\cite{raissi_physics-informed_2019}. Here, we add parentheses to clarify that PINNs technically refer to two types of problems: either treating the solution of a differential equation as an optimization problem, which can actually be done more efficiently by DNS, or the inverse problem of estimating parameters within the equation, as was done by~\citet{camporeale_data-driven_2022}. Otherwise, the term soft constraint should be used.

Another promising line of research for obtaining closure relations is equation discovery~\cite{camps-valls_discovering_2023}, which is a collection of methods that extract equations from data using symbolic or sparse regression. In sparse regression, a library of preselected expressions is fitted~\cite{brunton_discovering_2016,alves_data-driven_2022,ingelsten_data-driven_2025,donaghy_search_2023}. These methods have also been applied in conjunction with data augmentation~\cite{mcgrae-menge_embedding_2025}, such as applying Lorenz/Galilean boosts that enforce such invariance and improve the fidelity of the learned models. We would like to emphasize that methods such as physics-informed sparse regression~\cite{both_deepmod_2021}, Genetic Programming, and pre-trained transformers~\cite{landajuela_unified_2022} warrant more attention in plasma physics with regard to these types of problems. Naturally, as is the case with other forms of Machine Learning, these methods are prone to overfitting when presented with partial data and high expressivity (complexity of the expressions that can be fitted by the method). This is where intuition regarding physics-based closures~\cite{hunana_introductory_2019,hunana_introductory_2019-1} can be very useful in restricting the set of possibilities a priori. We firmly believe that progress in this field is possible by a careful combination of Machine Learning, high-performance computing, and theoretical considerations. 

%%%%%%%%%%%%%%%%%%%%%%%%%%%%%%%%%%%%%%%

%%%%%%%%%%%%%%%%%%%%%%%%%%%%%%%%%%%%%%%

\section{Conclusion}\label{sec:conclusion}

This work is the first application of non-local neural closure for the electron pressure tensor, achieved via a Fully Convolutional Neural Network (FCNN). Using a combination of statistical and physical fidelity diagnostics, such as pressure-strain and agyrotropy, we have demonstrated the generalization of this new closure from noisy (fewer particles per cell) Particle-in-Cell (PIC) simulations to more accurate (higher particle counts per cell) simulations. Pressure-strain diagnostics indicate that the closure accurately captures overall energy channels and certain local characteristics near the X-point of the reconnection site. This is promising, as we run PIC simulations for training data generation; however, with a higher number of particles per cell, simulations become prohibitively expensive to run in large quantities. Crucially, we demonstrate that FCNN significantly outperforms known closure relations, such as the previously used Multi-Layer Perceptron (MLP) or other double adiabatic-type models. We have addressed this problem in the context of Earth's magnetosheath decaying turbulence simulations, considering a specific set of physical parameters associated with large ion $\beta$ and moderate electron $\beta$. Future work will involve extending the validity of this closure to a broader set of parameters, 3D geometry, and coupling it to Reduced Order Models (ROMs), such as two-fluid and hybrid kinetic simulations. It will contribute to the development of efficient multi-scale models capable of probing larger domains of magnetospheric physics while accurately representing small-scale physics.

\begin{acknowledgments}
G.M. dedicates this work to the memory of Giovanni Lapenta, whose guidance, insight, and goodwill were essential.

We would like to thank Thierry Passot, Pierre-Henri, Silvio Cerri, Francesco Carella, Rony Keppens, Stefaan Poedts, and Maria Elena Innocenti for their valuable discussions and support.

G.M.\ has received funding from the European Union’s Horizon research and innovation program under the Marie Sk{\l}odowska-Curie grant agreement No 101148539. N. F. Oliveira Lopes and P. Dazzi would like to acknowledge the Research Foundation – Flanders (FWO), HELIOSKILL project grant (G0B9923N). 

The resources and services used in this work were provided by the VSC (Flemish Supercomputer Center), funded by the Research Foundation - Flanders (FWO) and the Flemish Government.

\end{acknowledgments}

\section*{Data Availability Statement}

The test datasets A and B1 used as test sets in this study are available in the Zenodo repository at \href{https://zenodo.org/records/17882782}{https://zenodo.org/records/17882782}.
The code is available at \href{https://github.com/georgemilosh/closure.git}{https://github.com/georgemilosh/closure.git}

\appendix

\section{Scale-to-scale analysis}\label{sec:scale2scale}

To investigate the behavior of energy conversion channels at different length scales, we employ the scale-filtering/coarse-graining method, widely used to analyze both magnetohydrodynamic \cite{yang_energy_2016,yang_compressibility_2017,arro_spatiotemporal_2025} and plasma \cite{yangEnergyTransferChannels2017,yangEnergyTransferPressure2017,matthaeus_pathways_2020,arro_spectral_2022} turbulence. We introduce a general filtering operation
\begin{equation}
\bar{f}_{\ell}^s(\boldsymbol{x}, t)=\int d^d r \; G_{\ell}(\boldsymbol{r}) f_s(\boldsymbol{x}+\boldsymbol{r}, t),
\end{equation}
where, for the remainder of the manuscript, we will only use the box-car filter following~\citet{matthaeus_turbulence_2021}. In addition, we will need the ``Favre-filtered'' (density-weighted filtered) quantities
(\ref{eq:mynum}),
\begin{equation}
\tilde{f}_{\ell}^s=\overline{\rho f_{\ell}^s} / \bar{\rho}_{\ell}^s
=\label{eq:mynum}
\end{equation}
Below, we present a scale-to-scale filtered version of the equations~\eqref{eq:energy_channels}, where we have removed the spatial transport terms by performing spatial {box} averaging $\langle, \rangle$.
\begin{subequations}
\label{eq:whole}
\begin{equation}
\partial_t \langle \widetilde{E}_s^f\rangle =-\langle \Pi_s^{V V} \rangle -\langle \Phi_s^{V T}\rangle -\langle \Lambda_s^{V b}\rangle,\label{subeq:1}
\end{equation}
\begin{eqnarray}
\partial_t \langle \bar{E}^m\rangle =-\sum_s \langle \Pi_s^{b b}\rangle +\sum_s \langle \Lambda_s^{V b}\rangle.\label{subeq:2}
\end{eqnarray}
\end{subequations}

Spatial averaging symbol will be omitted for the remainder of the manuscript. Here, the \emph{filtered fluid flow energy} is given by  
\begin{equation}
\widetilde{E}_s^{{f}} = \frac{1}{2} \bar{\rho}_s \tilde{V}_s^2 ,
\end{equation}
and the \emph{filtered electromagnetic energy} is  
\begin{equation}
\bar{E}^m = \frac{\overline{\boldsymbol{B}}^2 + \overline{\boldsymbol{E}}^2}{8\pi} .
\end{equation}

%The \emph{spatial transport terms} are defined as  
%\begin{equation}
%\boldsymbol{J}_s^V = \widetilde{E}_s^f \tilde{\boldsymbol{V}}_s 
%+ \overline{\rho}_s \tilde{\boldsymbol{\tau}}_s^V \cdot \tilde{\boldsymbol{V}}_s
%+ \overline{\boldsymbol{P}_s} \cdot \tilde{\boldsymbol{V}}_s ,
%\end{equation}
%and  
%\begin{equation}
%\boldsymbol{J}^b = \frac{c\,\overline{\boldsymbol{E}} \times \overline{\boldsymbol{B}}}{4\pi} .
%\end{equation}

The \emph{sub-grid-scale (SGS) flux of fluid flow energy across scales} due to nonlinearities is  
\begin{equation}
\Pi_s^{{VV}} = -\left(\bar{\rho}_s\,\tilde{\boldsymbol{\tau}}_s^V \cdot \nabla\right) \cdot \tilde{\boldsymbol{V}}_s 
- \frac{q_s}{c} \bar{n}_s\,\tilde{\boldsymbol{\tau}}_s^b \cdot \tilde{\boldsymbol{V}}_s ,
\end{equation}
where  
\begin{equation}
\tilde{\boldsymbol{\tau}}_s^V = \widetilde{\boldsymbol{V}_s \boldsymbol{V}_s} - \tilde{\boldsymbol{V}}_s \tilde{\boldsymbol{V}}_s , 
\quad  
\tilde{\boldsymbol{\tau}}_s^b = \widetilde{\boldsymbol{V}_s \times \boldsymbol{B}} - \tilde{\boldsymbol{V}}_s \times \tilde{\boldsymbol{B}} .
\end{equation}

The \emph{SGS flux of electromagnetic energy across scales} due to nonlinearities is  
\begin{equation}
\Pi_s^{bb} = -q_s \bar{n}_s \tilde{\boldsymbol{\tau}}_s^e \cdot \tilde{\boldsymbol{V}}_s , \quad \text{where} \quad \tilde{\boldsymbol{\tau}}_s^e = \tilde{\boldsymbol{E}} - \overline{\boldsymbol{E}} .
\end{equation}

The \emph{rate of conversion of flow energy into internal energy} is  
\begin{equation}
\Phi_s^{{V}T} = -\left(\overline{\boldsymbol{P}_s} \cdot \nabla\right) \cdot \tilde{\boldsymbol{V}}_s.
\end{equation}
The \emph{rate of conversion of fluid flow energy into electromagnetic energy} is  
\begin{equation}
\Lambda_s^{{V}b} = -q_s \bar{n}_s \widetilde{\boldsymbol{E}} \cdot \tilde{\boldsymbol{V}}_s ,
\end{equation}

\section{Datasplit 2}\label{sec:datasplit2}

In Table~\ref{tab:ablation2} we present results of study similar to Table~\ref{tab:R2_models} and Figure~\ref{fig:ablation_T2D10c1} but applied to a different datastplit, see Table~\ref{tab:sim_datasplit}. It appears that $R^2$ score for both diagonal and off-diagonal pressure tensor is consistent, which bolsters the robustness of the study. 

\begin{table}[h!]
\renewcommand{\arraystretch}{1.5}
\centering
\caption{\label{tab:ablation2}Ablation study on datasplit 2 (see Table~\ref{tab:sim_datasplit}): comparison between FCNN trained on different inputs. Model referred to as \emph{``default''} consists of $(n, \mathbf{v}_e, \mathbf{E}, \mathbf{B})$ inputs, \emph{``noE''} corresponds to $(n, \mathbf{v}_e, \mathbf{B})$, \emph{``Jtot''} corresponds to $(n, \mathbf{J}, \mathbf{E}, \mathbf{B})$, \emph{``JtotnoE''} corresponds to $(n, \mathbf{J}, \mathbf{B})$.} % Optional caption
\newcolumntype{C}[1]{>{\centering\arraybackslash}m{#1}}
\begin{tabular}{|l||C{1.3cm}|C{1.3cm}|C{1.3cm}|C{1.3cm}|}
\hline
\textbf{Model} & \textbf{\textit{default}} & \textbf{\textit{noE}} & \textbf{\textit{Jtot}} & \textbf{\textit{JtotnoE}} \\ 
\hline\hline

% Group 1
$p_\parallel$ & 0.78 & {0.79} & 0.78 & {0.81} \\
$p_\perp$     & 0.84 & 0.82 & 0.83 & 0.84 \\ 
\hline

% Group 2
$P_{xx}$        & 0.81 & 0.80 & 0.80 & {0.82} \\
$P_{yy}$        & 0.82 & 0.80 & 0.81 & {0.83} \\
$P_{zz}$        & 0.77 & 0.77 & 0.77 & {0.80} \\ 
\hline

% Group 3
$P_{xy}$        & 0.42 & {0.37} & 0.42 & {0.37} \\
$P_{xz}$        & 0.37 & {0.33} & 0.38 & 0.35 \\
$P_{yz}$        & 0.37 & {0.34} & 0.38 & 0.36 \\ 
\hline

\end{tabular}
\end{table}

\section{Galilean invariance}\label{ref:galilean}

\begin{figure*}[ht!]
\centering
\begin{tabular}{@{}c@{}c@{}c@{}} % The @{} removes inter-column padding

% --- ROW 1 ---
% Each panel is a `subfigure` environment. This is necessary for the \label to work.
% The width is set to a third of the text width to make them fit in a 3-column grid.
%\vspace{-.5cm}

\begin{subfigure}{0.45\textwidth}
  % Replace 'example-image-a.png' with your actual PNG file.
  % The [width=\linewidth] option scales the image to the width of the subfigure.
  \begin{overpic}[width=\linewidth]{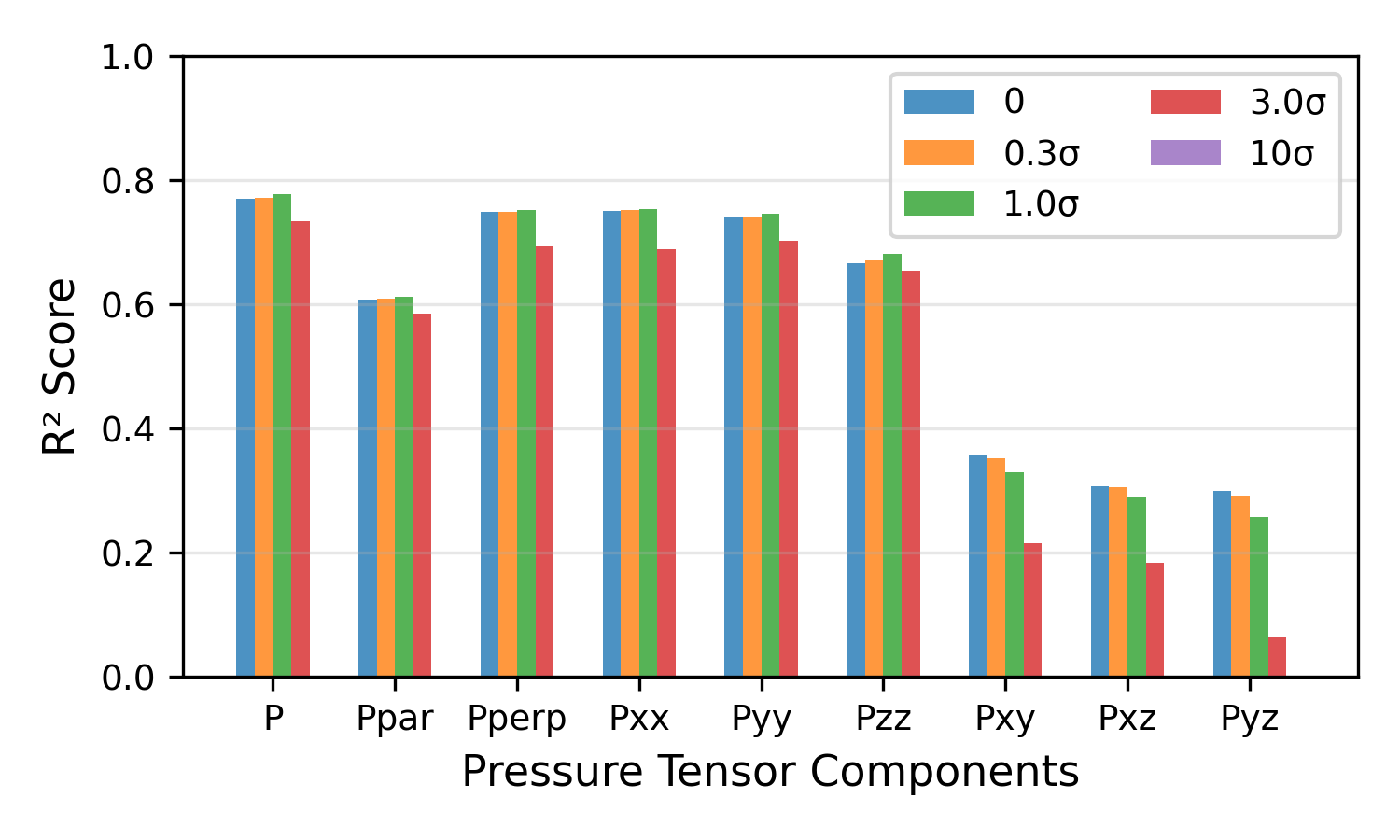}
    % \put(x,y){text}: Places 'text' at coordinates (x,y).
    % The coordinates are percentages of the image width and height.
    % (5,88) means 5% from the left and 88% from the bottom (i.e., top-left).
    % \color{white} makes the text white, which is good for dark backgrounds.
    \put(0,50){\panellabel{a}[black]} 
  \end{overpic}
  \phantomcaption % Empty caption, needed for the subfigure counter and label
  \label{fig:galilean_x}
\end{subfigure}&
\begin{subfigure}{0.45\textwidth}
  \begin{overpic}[width=\linewidth]{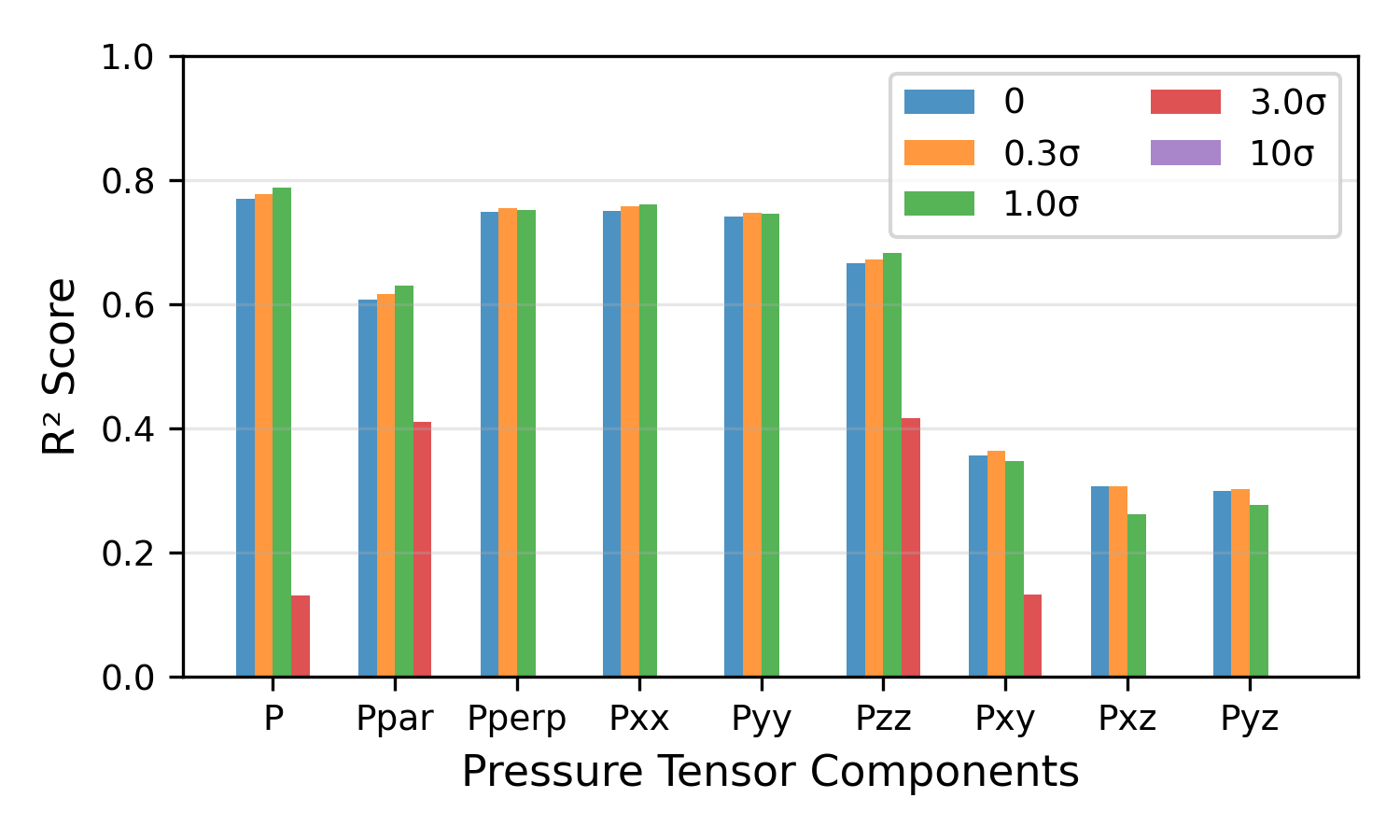}
    \put(0,50){\panellabel{b}[black]} 
  \end{overpic}
  \phantomcaption
  \label{fig:galiliean_z}
\end{subfigure}\\

\begin{subfigure}{0.45\textwidth}
  % Replace 'example-image-a.png' with your actual PNG file.
  % The [width=\linewidth] option scales the image to the width of the subfigure.
  \begin{overpic}[width=\linewidth]{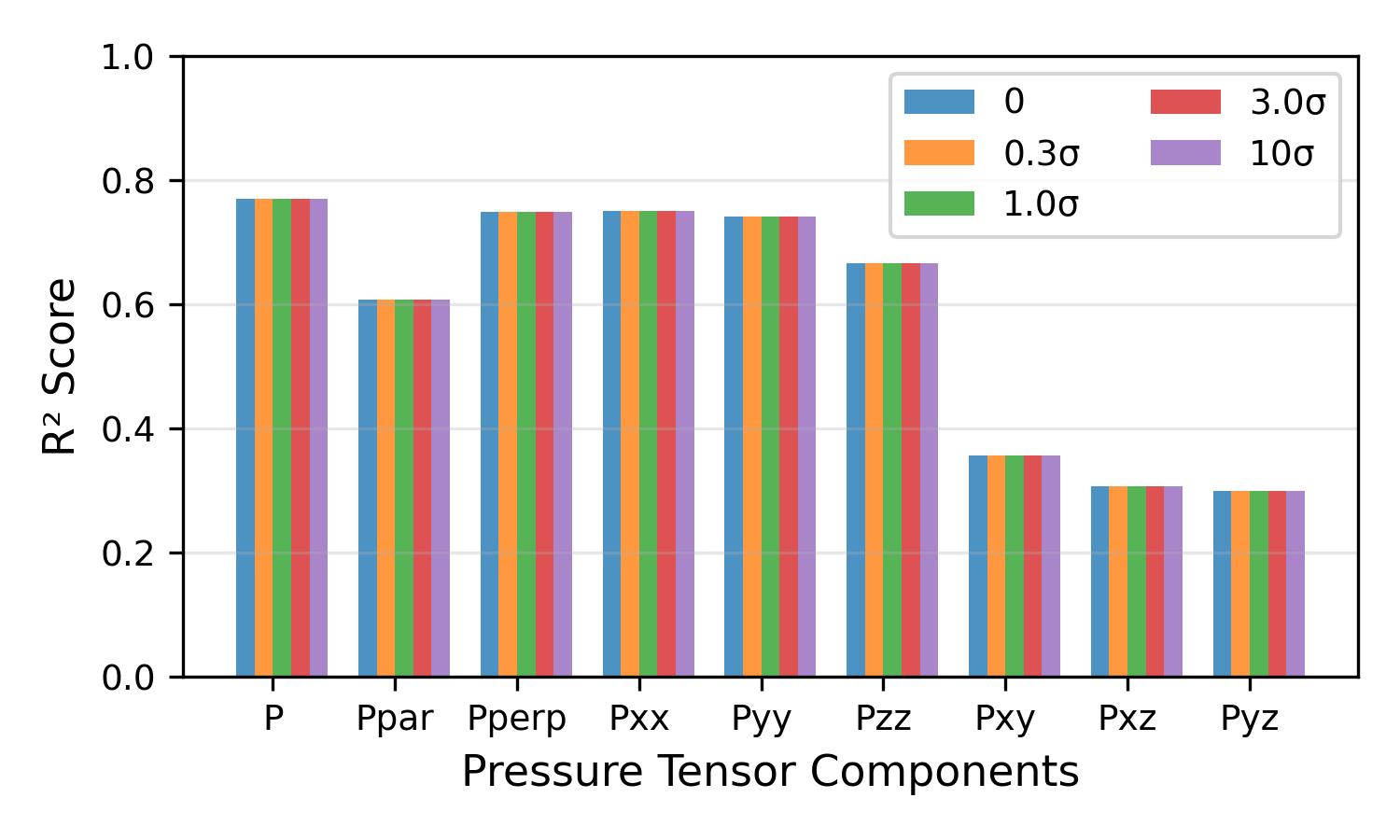}
    % \put(x,y){text}: Places 'text' at coordinates (x,y).
    % The coordinates are percentages of the image width and height.
    % (5,88) means 5% from the left and 88% from the bottom (i.e., top-left).
    % \color{white} makes the text white, which is good for dark backgrounds.
    \put(0,50){\panellabel{c}[black]} 
  \end{overpic}
  \phantomcaption % Empty caption, needed for the subfigure counter and label
  \label{fig:galilean_x_fix}
\end{subfigure}&
\begin{subfigure}{0.45\textwidth}
  \begin{overpic}[width=\linewidth]{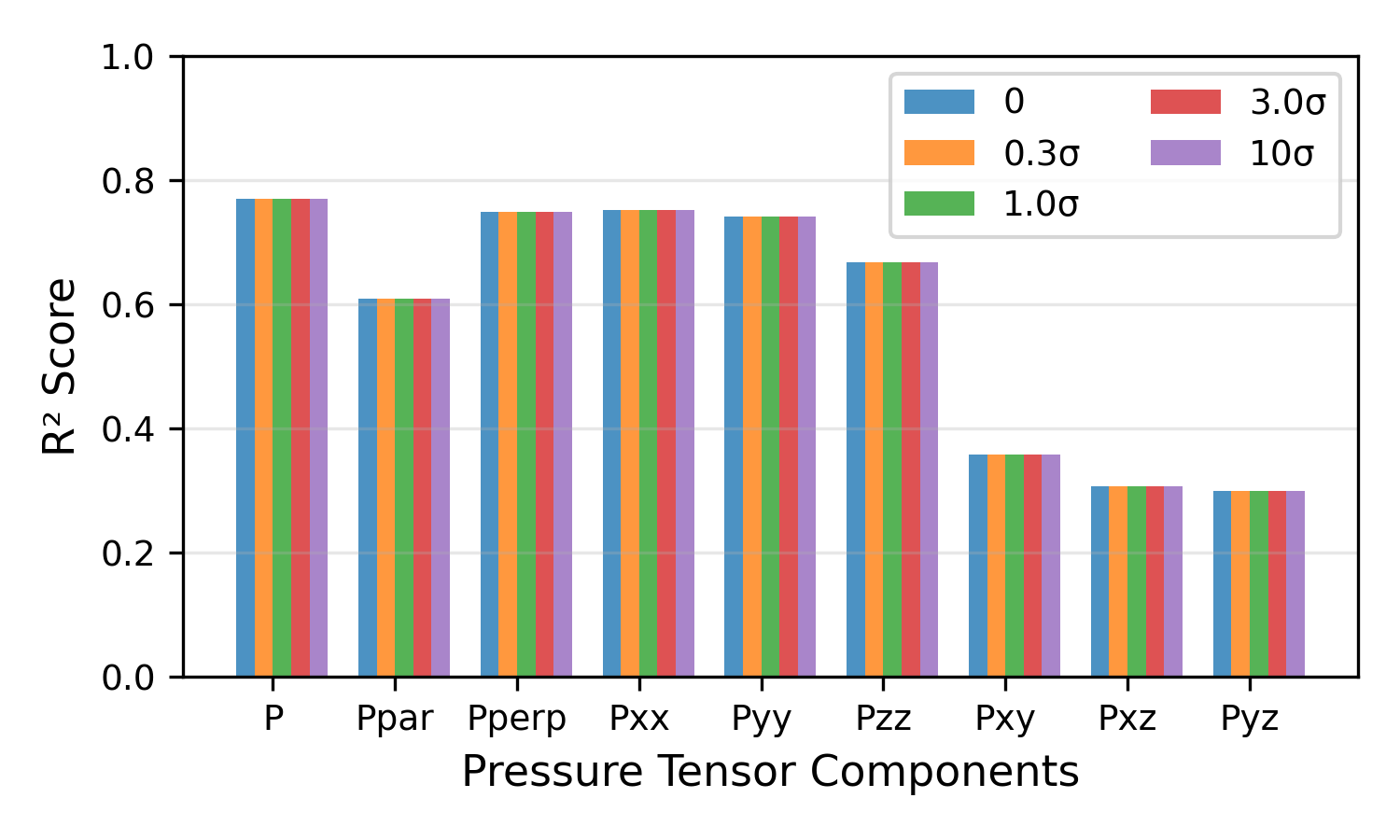}
    \put(0,50){\panellabel{d}[black]} 
  \end{overpic}
  \phantomcaption
  \label{fig:galiliean_z_fix}
\end{subfigure}
\end{tabular}
%/volume1/scratch/georgem/closure/models/peppe/sigma0_haydn/FCNN/PnoE/Galilean.ipynb
\caption{{Galilean invariance test over 4 different values of boosts corresponding to $v/c = \{0, 0.005, 0.02, 0.05\}$ (a) Applying $\mathbf{v} = v\,\hat{\mathbf{x}}$ (b) Applying $\mathbf{v} = v\,\hat{\mathbf{z}}$. (c) Applying $\mathbf{v} = v\,\hat{\mathbf{x}}$ and subtracting the mean $v_x$ from the input to the neural network (d), applying $\mathbf{v} = v\,\hat{\mathbf{z}}$ and subtracting the mean $v_z$ from the input to the neural network.  }}
\label{fig:galilean}
\end{figure*}

{In what follows, we apply the Galilean boosts $\mathbf{x}^\prime = \mathbf{x} -\mathbf{v}t$, which are also associated with the transformation of the electromagnetic field that under $\mathbf{v}\ll c$}
\begin{equation}
    {\mathbf{E}^\prime = \mathbf{E} + \mathbf{v} \times \mathbf{B}}
\end{equation}
and
\begin{equation}\label{eq:Bgalilean}
    {\mathbf{B}^\prime = \mathbf{B} - \frac{1}{c^2}\mathbf{v} \times \mathbf{E}}
\end{equation}
{The primes denote quantities in the boosted reference frame. The values are provided as inputs. These transformations are applied to the neural network's input to predict pressure. We plot the determination scores for a range of values of boosts $v/c = \{0, 0.005, 0.02, 0.05, 0.2$ in Figure~\ref{fig:galilean}. The values are chosen to correspond to fractions of variance of the corresponding component of the velocity $\sigma_x:= \sqrt{\langle v_x^2 \rangle}$ in Figure~\ref{fig:galilean_x} and $\sigma_z:= \sqrt{\langle v_z^2 \rangle}$ in Figure~\ref{fig:galiliean_z}. From Figure~\ref{fig:galilean_x} we see that the network is capable of reconstructing the diagonal pressure tensor components up to $3\sigma$. From Figure~\ref{fig:galiliean_z} we see that performance drops to 1 $\sigma$ above which the network fails. This is because as we increase $\mathbf{v}$, the network is supplied with a larger distribution of velocity that it has not seen during training. One way to address this problem is to augment the dataset by applying random Galilean boosts during training. This is similar to the recent work~\cite{mcgrae-menge_embedding_2025} that embeds Lorentz invariance via data augmentation. However, for our purposes, we can show that this is not necessary since we can simply augment the pressure model~\eqref{eq:pressure-model} using the following anzats:  }
\begin{equation}\label{eq:pressure-model-galilean}
{\mathbf{P} = \mathbf{P}_\theta (n, \mathbf{V}_e - \langle \mathbf{V}_e \rangle, \mathbf{E}, \mathbf{B}),}
\end{equation}
{where $\langle \rangle $ operation implies spatial averaging. This is meant to undo the main contribution of the Galilean boost, and no additional training is needed, since during training $\langle \mathbf{V}_e \rangle\approx 0$. This trivial operation yields updated Figures for both the $v_x$ (Figure~\ref{fig:galilean_x_fix}) and $v_z$ boosts (Figure~\ref{fig:galiliean_z_fix}) and we can see that the performance of closure is independent of the boost value because the changes in $B$ field (equation~\eqref{eq:Bgalilean}), which are also inputs to equation~\eqref{eq:pressure-model-galilean}, are subdominant. }
%\nocite{*}
\bibliography{references1}

\end{document}